\documentclass[10pt,english,notitlepage,a4paper,english,superscriptaddress,floatfix,longbibliography,pre]{revtex4-1}
\usepackage[T1]{fontenc}
\usepackage{babel}
\usepackage{amsmath}
\usepackage{amssymb}
\usepackage{graphicx}
\usepackage{bm}
\usepackage{ulem}
\usepackage[unicode=true,pdfusetitle,
 bookmarks=true,bookmarksnumbered=false,bookmarksopen=false,
 breaklinks=false,pdfborder={0 0 1},backref=false,colorlinks=true]
 {hyperref}
\usepackage{comment}
\usepackage[percent]{overpic}
\makeatletter

\newcommand{\N}[0]{\bar{N}\rule{0pt}{3.5mm}}
\newcommand{\figpanel}[1]{(\textbf{\lowercase{#1}})}
\usepackage{xcolor}
\definecolor{myblue}{RGB}{12, 12, 158}
\definecolor{myred}{RGB}{158, 19, 22}
\definecolor{myorange}{RGB}{245, 150, 12}
\definecolor{mygreen}{RGB}{26, 148, 49}
\definecolor{Prune}{RGB}{99,0,60}
\definecolor{Purple}{RGB}{75, 0, 130}
\definecolor{Pink}{RGB}{255, 105, 180}
\definecolor{deepskyblue}{RGB}{0, 191,255}
\definecolor{limegreen}{RGB}{50, 205, 50}
\definecolor{crimson}{rgb}{0.86, 0.08, 0.24}

\definecolor{blue(ncs)}{rgb}{0.0, 0.53, 0.74}

\newcommand{\repidx}[1]{12}
\makeatother

\begin{document}
\title{Thermodynamics of bidirectional associative memories}
\author{Adriano Barra}
\affiliation{Dipartimento di Matematica e Fisica, Università del Salento, Campus Ecotekne, 73100 Lecce, Italy.}
\affiliation{Istituto Nazionale di Fisica Nucleare, Sezione di Lecce, Campus Ecotekne, 73100 Lecce, Italy.}

\author{Giovanni Catania}
\email{gcatania@ucm.es}
\affiliation{Departamento de Física Teórica, Universidad Complutense de Madrid, 28040
Madrid, Spain.}
\author{Aurélien Decelle}
\affiliation{Departamento de Física Teórica, Universidad Complutense de Madrid, 28040
Madrid, Spain.}
\affiliation{Université Paris-Saclay, CNRS, INRIA Tau team, LISN, 91190 Gif-sur-Yvette,
France.}
\author{Beatriz Seoane}
\affiliation{Departamento de Física Teórica, Universidad Complutense de Madrid, 28040
Madrid, Spain.}
\begin{abstract}
In this paper we investigate the equilibrium properties of bidirectional associative memories (BAMs). Introduced by Kosko in 1988 as a generalization of the Hopfield model to a bipartite structure, the simplest architecture is defined by two layers of neurons, with synaptic connections only between units of different layers: even without internal connections within each layer, information storage and retrieval are still possible through the reverberation of neural activities passing from one layer to another. We characterize the computational capabilities of a stochastic extension of this model in the thermodynamic limit, by applying rigorous techniques from statistical physics. A detailed picture of the phase diagram at the replica symmetric level is provided, both at finite temperature and in the noiseless regimes.  Also for the latter, the critical load is further investigated up to one step of replica symmetry breaking. An analytical and numerical inspection of the transition curves (namely critical lines splitting the various modes of operation of the machine) is carried out as the control parameters - noise, load and asymmetry between the two layer sizes - are tuned. In particular, with a finite asymmetry between the two layers, it is shown how the BAM can store information more efficiently than the Hopfield model by requiring less parameters to encode a fixed number of patterns. Comparisons are made with numerical simulations of neural dynamics. Finally, a low-load analysis is carried out to explain the retrieval mechanism in the BAM by analogy with two interacting Hopfield models. A potential equivalence with two coupled Restricted Boltmzann Machines is also discussed.

\end{abstract}

\maketitle

\section{Introduction}

Since the original work of Amit, Gutfreund, and Sompolinsky (AGS)~\citep{amit_storing_1985}
who first developed a mean-field theory for the Hopfield model of neural networks in terms of spin glasses, several extensions and generalizations have been made. As for dilution, although synaptic interactions in the Hopfield model are fully connected, it is possible to study artificial networks on sparse topologies, for example, by considering synaptic matrices arranged in random graphs, hierarchical structures, small-world, or scale-free architectures \citep{castillo2004little,agliari_analogue_2012,castillo_coolen_hopfield_scalefree,guerra-hierarchical}.
Another possibility is to consider sparsity in the pattern components, leading
to equilibrium regimes where the retrieval of composite information is
possible~\citep{agliari_multitasking_2012}. In addition, extensions to different
support for the spin variables and/or the patterns (e.g. real-valued
spins and/or pattern components) have been deeply analyzed \citep{bolle_spherical_2003,leuzzi_quantitative_2022,agliari_analogue_2012,barra_phase_2018}.

One of the most studied generalizations of the Hopfield model is a bipartite structure, where the network is divided into two layers of neurons and connections are allowed only between neurons of different layers. The most popular application of this type of topology is the so-called Restricted Boltzmann Machine (RBM) \citep{smolensky1986information,shimagaki2019selection,aurelien_decelle_restricted_2021}, which is widely used in computer science~\citep{smolensky1986information,hinton2002training,hinton2006reducing} (e.g., as fundamental units of modern deep architectures) and can be considered a prototype for machine learning models~\citep{hinton2006reducing}. The Hopfield model was proposed as a toy model for neurophysiology that accounts for biological learning à la Hebb, and yet strong similarities between the information processing mechanisms of these two types of models have been demonstrated by statistical mechanical studies~\citep{barra_equivalence_2012,Leonelli-NN2021,shimagaki2019selection,aurelien_decelle_restricted_2021}, whose modus operandi will therefore be the methodological leitmotif of the present research.

Two main classes of reward emerge from the duality between neural architectures used in machine learning (e.g., RBMs) and biologically inspired neural networks (e.g., Hebbian models). The first is explainability, since by mapping the information processing capabilities of RBMs to those that naturally emerge in the Hopfield network, we can better understand the hidden decision-making mechanisms the machine follows when it performs, for example, denoising or pattern recognition: this can be particularly welcome in the field of Explainable Artificial Intelligence~\citep{alemanno_supervised_2022}. Moreover, statistical mechanics eventually provides phase diagrams, i.e. it reveals regions in the space of control parameters (e.g., noise, load, etc.) where the machine has certain information storage capabilities, and regions where these properties are lost. This, in turn, is particularly welcome in the field of sustainable Artificial Intelligence, since knowledge of the phase diagram allows the machine to be brought into the optimal operational setting, which can lead to significant savings in training costs. In particular, the equivalence between the Hopfield model and certain classes of Boltzmann Machines \citep{barra_equivalence_2012,shimagaki2019selection,aurelien_decelle_restricted_2021}
has shown a strong connection between the retrieval properties of biological neural networks and the performance of machine learning algorithms in terms of their ability to correctly describe the empirical distribution of a dataset and/or eventually overfit it \citep{barra_equivalence_2012}; furthermore, the retrieval properties of the Hopfield model can be used to develop new training procedures for RBMs~\citep{Pozas_Kerstjens_2021}.
This means that the properties of biological and artificial information processing networks can be analyzed and exploited using similar techniques.

The bidirectional associative memory (BAM) we study in this paper was introduced by Kosko in \citep{kosko_bidirectional_1988} as an attempt to overcome the lack of internal organization of information in the original Hopfield model and to account for structured retrieval of patterns. It is another generalization of a neural network based on a bipartite topology: in this machine, as in RBM settings, there are only interactions between units of different layers. However, while RBMs have two layers covering rather different computational roles (the input layer is tipically provided with the datasets, e.g. noisy patterns to be recognized and/or classified, and the hidden layer derives correlations in the visible variables from the supplied information, enabling pattern recognition and/or noise reduction) and the weight matrix has no particular biological significance, in the BAM architecture both layers are expected to provide pattern information (one pattern per layer, as an attempt to retrieve pairs of patterns rather than individual ones) and their synaptic matrix strongly resembles the original Hebb proposal. However, the operating principle of the BAM differs from that of the Hopfield model because in the BAM retrieval, information is passed from one layer to another according to appropriate dynamic rules: this mechanism is referred to in the literature as \textit{reverberation} of information ~\citep{kosko_bidirectional_1988}.
The BAM was analyzed using statistical physics techniques in previous works ~\citep{englisch_bm_1995,kurchan_statistical_1994}, but only in the noiseless case (i.e. at $T=0$) since the goal was to determine the storage properties in the saturation regime. Further analysis has been carried out in \citep{tanaka2000capacity} using the replica trick at the replica symmetric level. In this paper, we perform an extensive analysis of the equilibrium properties of the BAM: we characterize the phase diagram analytically and numerically, and provide an extension of the Hopfield AGS mean field theory to the BAM. Analytical calculations 
are obtained independently using the Guerra interpolation scheme~\citep{guerra_course_2006} and the replica method~\citep{mezard1987spin}, both at the replica symmetric and with $1$ step of replica symmetry breaking. Finally, the BAM has recently become topical again in the context of machine learning~\citep{Pozas_Kerstjens_2021}. In this context, a detailed study of the phase diagram of the BAM in the presence of an external noise (i.e., at $T\ne0$) and in terms of the asymmetry between the two layers is necessary to clarify under which circumstances retrieval of information is possible or not, a necessary first step to analyze the pattern extraction process with RBMs.

The paper is organized as follows: Sect.~\ref{sec:Model} introduces the model and the notation used in the rest of the paper. Sect.
\ref{sec:Equilibrium-analysishigh-load} discusses the equilibrium behavior of the model in the thermodynamic limit by analyzing the free energy density computed under the assumption of replica symmetry (RS): a detailed discussion of the phase diagram of the model is given both in the presence of noise and in the noiseless limit; furthermore, Section \ref{sec:1RSB} presents a more peculiar analysis obtained using a 1-step Replica Symmetry Breaking ansatz. Section \ref{sec:numericalsim} reports some numerical simulations used to test theoretical results. Section \ref{sec:Pattern-retrieval-in}
provides a simple justification of how pattern retrieval is attained in the BAM by interpolating the model with two independent Hopfield models (solely in the low-load regime for the sake of simplicity). Finally, Sect.
\ref{sec:Conclusion} summarizes our results and presents future directions to be explored. Calculation details are relegated to the Appendix, where a formal equivalence between the BAM and two coupled RBMs is further discussed.

\section{Model\label{sec:Model}}

\begin{figure}
  \includegraphics[width=0.5\textwidth]{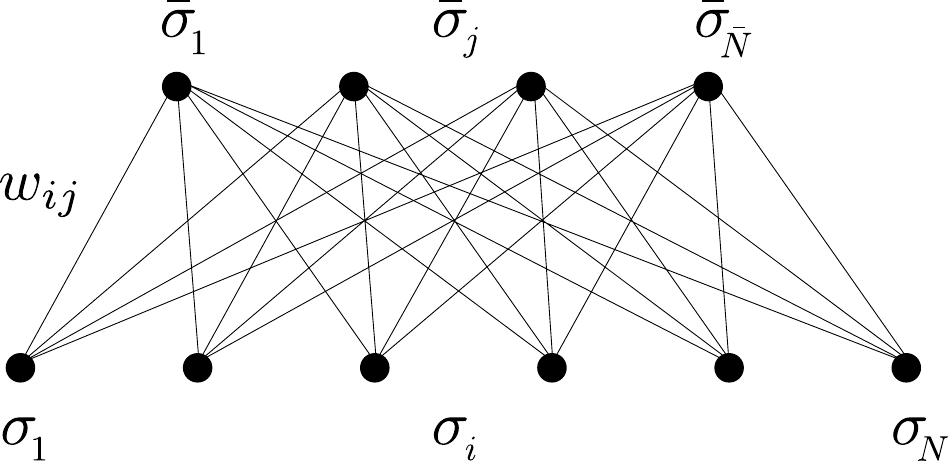}
  \caption{Schematic representation of a BAM with $N=6$ and ${\N}=4$. The (binary) neurons are represented by black circles; each solid line defines a synaptic weight connecting two neurons of different layers.\label{fig:BAM_scheme}}
\end{figure}

The BAM model is defined by two
sets of neurons, embedded in two vectors $\boldsymbol{\sigma}$ and
$\bar{\boldsymbol{\sigma}}$ for layer $1$ and $2$, respectively.
Each unit in one layer interacts with all the units in the other
layer, so that the interaction topology is a bipartite graph; for a schematic representation of the model, see Figure \ref{fig:BAM_scheme}.  Layer
$1$ (resp. $2$) has $N$ (resp. ${\N}$) neurons, called - in the first (resp. second layer) as $\sigma_{i}$
(resp $\bar{\sigma}_{j}$) with $i\in\left\{ 1,\ldots,N\right\} $
(resp. $j\in\{ 1,\ldots,{\N}\} $) \footnote{In the whole discussion, we will always indicate with an overbar $\bar{\cdot}$ quantities referring to layer $2$.}. Each neuron is hereafter
considered as a binary variable (i.e. an Ising spin), namely $\sigma_{i},\bar{\sigma}_{j}\in\left\{ -1,1\right\}$,
where the state $+1$ is typically associated with a firing state for
the neuron, while $-1$ means the neuron is silent. The BAM Hamiltonian (or cost-function) is given by:\begin{equation}
H\left(\boldsymbol{\sigma},\bar{\boldsymbol{\sigma}}\right)=-\sum_{i=1}^{N} \sum_{j=1}^{\N} w_{ij}\sigma_{i}\bar{\sigma}_{j},\label{eq:hamiltonian_starting}
\end{equation}
where $w_{ij}$ represents the synaptic interaction between node $i$
in the first layer and the node $j$ in the second layer. The interaction matrix $\boldsymbol{W}=\left\{ w_{ij}\right\} _{i=1,\ldots,N}^{j=1,\ldots,{\N}}$
is constructed in such a way that the usual Hebb's rule pattern storage is generalized to such a bipartite structure. We therefore define
two sets of $K$ patterns (one per layer), denoted with $\boldsymbol{\xi}^{\mu}$
and $\bar{\boldsymbol{\xi}}^{\mu}$, with $\mu \in \left \{1, \ldots, K\right \}$. Each pattern has a dimension
compatible with the corresponding layer of neurons, so that $\boldsymbol{\xi}^{\mu}, \bar{\boldsymbol{\xi}}^{\mu}$ are vectors with $N$ and ${\N}$ components, respectively, for each $\mu$. The synaptic
weight matrix is constructed by using the following generalized Hebb's rule:
\begin{align}
w_{ij} & =\frac{1}{\sqrt{N\N}} \sum_{\mu=1}^{K}\xi_{i}^{\mu}\bar{\xi}_{j}^{\mu}.\label{eq:coupling_matrix}
\end{align}
The scale factor in the weight matrix is chosen so to have a non-trivial free energy density in the thermodynamic limit where $N,{\N},L:=\sqrt{N{\N}}\to\infty$.

To inspect the ability of the BAM  to handle pairs of patterns, given the bipartite structure of the network and the absence of intra-synaptic interactions, it is useful to check that retrieval is feasible when both layers have a non-zero overlap with a given pattern. In what follows, we assume that the patterns are drawn independently with i.i.d. components at each layer (see the next section for more details on the distribution from which they are drawn). Under appropriate initial and external conditions (the latter depend on the number of patterns $K$), a neural dynamics starting with one of the layers close enough to one of the $K$ patterns will eventually reach a fixed point with a certain nonzero overlap on both layers: In the noiseless regime, the attractors of the dynamics - in pairs $(\boldsymbol{\xi}^{\mu},\bar{\boldsymbol{\xi}}^{\mu})$ - satisfy the following steepest descent dynamical equations:

\begin{align}
\sigma_{i}^{t+1} &=\text{sign}\left(\sum_{j}w_{ij}\bar{\sigma}^{t}_{j}\right) \qquad \longrightarrow \qquad \xi_{i}^{\mu}=\text{sign}\left(\sum_{j}w_{ij}\bar{\xi}_{j}^{\mu}\right) \label{eq:attractors1}\\
\bar{\sigma}_{j}^{t+1}& =\text{sign}\left(\sum_{i}w_{ij}\sigma^{t}_{i}\right) \, \qquad \longrightarrow \qquad  \bar{\xi}_{j}^{\mu}=\text{sign}\left(\sum_{i}w_{ij}\xi_{i}^{\mu}\right). \label{eq:attractors2}   
\end{align}
where the right equations highlight how pattern pairs $(\boldsymbol{\xi}^{\mu},\bar{\boldsymbol{\xi}}^{\mu})$ are fixed points of such a dynamics. Here we used the orthogonality between patterns, which holds in the thermodynamic limit, namely
\begin{equation}
    \frac{1}{N}\sum_i \xi_i^{\mu} \xi_i^{\nu} = \delta_{\mu, \nu} + O\left(\frac{1}{\sqrt{N}} \right) \qquad \text{and} \qquad \frac{1}{\N}\sum_j \bar{\xi}_j^{\mu} \bar{\xi}_j^{\nu} = \delta_{\mu, \nu} + O\left(\frac{1}{\sqrt{\N}} \right).
\end{equation}
It is easy to check that pattern pairs (i.e. the attractors of Eqs. \eqref{eq:attractors1}-\eqref{eq:attractors2}) are extremal points of the energy function,
by writing it in the following form:
\begin{align}
H\left(\boldsymbol{\sigma},\bar{\boldsymbol{\sigma}}\right)  =-\sum_{i,j}w_{ij}\sigma_{i}\bar{\sigma}_{j} =-\frac{1}{L}\sum_{\mu=1}^{K}\left(\sum_{i=1}^{N}\xi_{i}^{\mu}\sigma_{i}\right)\left(\sum_{j=1}^{{\N}}\bar{\xi}_{j}^{\mu}\bar{\sigma}_{j}\right) 
 = -L\sum_{\mu=1}^{K} m_\mu(\bm{\sigma}) \bar{m}_\mu(\bm{\bar{\sigma}}),
\end{align}
where we introduced, as order parameters of the theory, the Mattis overlap $m_{\mu}(\bm{\sigma})$, $\bar{m}_{\mu}(\bm{\bar{\sigma}})$ of the configurations $\left( \bm{\sigma},\bar{\bm{\sigma}} \right) $ with the pattern pair $\left( \bm{\xi}^{\mu},\bar{\bm{\xi}}^{\mu} \right)$ defined as
\begin{equation}
    m_{\mu}(\bm{\sigma}) = \frac{1}{N} \sum_i^{N} \xi_i^\mu \sigma_i \;\;\; \text{and } \;\; \bar{m}_{\mu}(\bm{\bar{\sigma}}
    ) = \frac{1}{{\N}} \sum_j^{\N} \bar{\xi}_j^\mu \bar{\sigma}_j.\label{eq:mattis_overlap_starting}
\end{equation}
\section{Equilibrium analysis in the high-load regime\label{sec:Equilibrium-analysishigh-load}}
\subsection{Basic setting\label{subsec:basicsetting}}
In this section, we discuss how to characterize the phase diagram of the BAM using techniques from statistical physics. We begin by defining the partition function at a fixed realization of the pattern components, at a given inverse temperature $\beta \in \mathbb{R}^{+}$:
\begin{align}
Z\left(\beta,\left\{ \left(\boldsymbol{\xi}^{\mu},\bar{\boldsymbol{\xi}}^{\mu}\right)\right\} _{\mu=1}^{K}\right) & =\sum_{\boldsymbol{\sigma}}\sum_{\bar{\boldsymbol{\sigma}}}\exp\left[\frac{\beta}{L}\sum_{\mu=1}^{K}\left(\sum_{i=1}^{N}\xi_{i}^{\mu}\sigma_{i}\right)\left(\sum_{j=1}^{{\N}}\bar{\xi}_{j}^{\mu}\bar{\sigma}_{j}\right)\right].\label{eq:partfunctionBAM}
\end{align}
The temperature $T=\beta^{-1}$ plays the role of an external noise
affecting the neural dynamics. Patterns are quenched with respect to the timescale of thermal fluctuations in the (physiological) neural dynamics, and are treated, as standard in spin-glass models, as quenched disorder
to be averaged over. Moreover, we assume that synaptic couplings already exist in the form \eqref{eq:coupling_matrix}, therefore no learning process is considered. The physical quantity containing all the information
about the equilibrium behaviour of the model is the quenched free
energy density
\begin{align}
f\left(\beta,\alpha,\gamma\right) & =-\frac{1}{\beta} \lim_{L\to\infty} \frac{1}{L} \mathbb{\mathbb{E}}_{\boldsymbol{\xi},\bar{\boldsymbol{\xi}}}\log Z\left(\beta,\left\{ \left(\boldsymbol{\xi}^{\mu},\bar{\boldsymbol{\xi}}^{\mu}\right)\right\} _{\mu=1}^{K}\right),\label{eq:quenched_p_starting}
\end{align}
where the operator $\mathbb{E}$ denotes the expectation value w.r.t. the patterns' distribution. In the thermodynamic limit, the free energy density will depend on three control parameters, i.e. the thermal noise $\beta$, the {\it network load} $\alpha$,
\begin{align}
\alpha & \equiv\frac{K}{L}\label{eq:alpha},
\end{align}
that reflects the ratio between the number of patterns embedded in the network
and the relevant system size of the model $L$, and the {\it asymmetry}  between layers (or the {\it shape} parameter),
\begin{align}
\gamma & \equiv\sqrt{\frac{N}{{\N}}}\,,\label{eq:gamma}
\end{align}
quantifying the ratio between the two layer sizes \footnote{The square root in Eq. \eqref{eq:gamma} is chosen for notation convenience}.

There are two main computational regimes under which this model can be analyzed: a first one in which the number of patterns is finite with respect to $L$ - or growing up to $\sim \log L$ - so that $\lim_{L\to\infty}\alpha=0$, called {\it low-load}
regime (whose treatment is mathematically simpler), and the more challenging {\it high-load} regime, in which the number of patterns also scales linearly with $L$, so that $\alpha \in \mathbb{R}^{+}$ even in the thermodynamic limit (and techniques from statistical mechanics of spin-glasses need to be introduced in the mathematical treatment). In both cases, we let both layer sizes go to infinity, while $\gamma$ remains finite. The low-load scenario was  analyzed in \citep{kurchan_statistical_1994}, where it was proven that a second-order phase transition splits a paramagnetic phase (with no retrieval) and a ferromagnetic phase, where retrieval spontaneously emerges. In this regime, when considering the retrieval of only one pair of patterns, the BAM shows a qualitatively similar behavior to the (bipartite) Curie-Weiss model \citep{bipartitecurieweiss_contuccigallo,Collet2014}.
We will come back to the low-load scenario in Sect.
\ref{sec:Pattern-retrieval-in}, where a simple justification of how the BAM achieves pattern-pair retrieval will be discussed. The high-load regime was also explored in Ref.~\citep{kurchan_statistical_1994}, but results were provided only at
zero-temperature. The calculations in the high-load regime are more delicate because they require a more careful treatment of the extensive noisy term in the Hamiltonian. 
The expectation value in Eq.~\eqref{eq:quenched_p_starting} is computed w.r.t. all the pattern
pairs $\left(\boldsymbol{\xi}^{\mu},\bar{\boldsymbol{\xi}}^{\mu}\right)$.
To describe the retrieval properties, we now divide the set of pattern pairs into two subsets: the first consists of a finite
(or up to $\sim\log L$) number of binary patterns, indexed by $\mu=1,\ldots,l$;
the remaining (extensive in the high-load regime) patterns are drawn from i.i.d. Gaussian for each component for $\mu=l+1,\ldots,K$. The first set encodes for pattern pairs that can be retrieved, i.e. those having eventually a non-zero overlap with spin configurations, while the other extensive set will act as a quenched noise w.r.t. the former. As a consequence, the Hamiltonian is split into a \textit{signal} and a \textit{noise} term as well:
\begin{equation}
H\left(\boldsymbol{\sigma},\bar{\boldsymbol{\sigma}}\right)=-\frac{1}{L}\sum_{\mu=1}^{l}\left(\sum_{i=1}^{N}\xi_{i}^{\mu}\sigma_{i}\right)\left(\sum_{j=1}^{{\N}}\bar{\xi}_{j}^{\mu}\bar{\sigma}_{j}\right)-\frac{1}{L}\sum_{\mu=l+1}^{K}\left(\sum_{i=1}^{N}\xi_{i}^{\mu}\sigma_{i}\right)\left(\sum_{j=1}^{{\N}}\bar{\xi}_{j}^{\mu}\bar{\sigma}_{j}\right).\label{eq:hamiltonian_separatedpatterns}
\end{equation}

The reason behind this choice generalizes standard arguments for
the Hopfield model with binary neurons, where information retrieval is optimally achieved in the saturation regime (i.e. at finite load $\alpha >0 $) when the patterns to be recalled are binary-valued, see for instance Refs. \citep{agliari_neural_2017,barra_phase_2017,barra_phase_2018}. On the other hand, the extensive part of the pattern set can be treated in differet ways, according to standard universality arguments about quenched disorder in spin-glass models~\citep{carmona_universality_2006,genovese_universality_2012}.
\paragraph{Decoupling:}

In fully connected spin glasses with quadratic interactions, the Boltzmann weight is usually simplified by introducing a Hubbard-Stratonovich integral transformation. The two-layer structure of the BAM instead requires a different decoupling transformation. Following standard approaches used to analyze bipartite models (such as the bipartite SK model \citep{hartnett_replica_2018} or Restricted Boltzmann machines \citep{decelle_spectral_2017,decelle_thermodynamics_2018}),
an integral transformation involving a pair of complex conjugate variables $\left(z,z^{\dagger}\right)$ is a more convenient way to decouple the interacting term in the Hamiltonian:
\begin{equation}
\exp\left[\beta x\bar{x}\right]=\frac{\beta}{\pi}\int dzdz^{\dagger}\exp\left[-\beta zz^{\dagger}+\beta zx+\beta z^{\dagger}\bar{x}\right].\label{eq:integral_transform_complex_generic}
\end{equation}
The above identity can be easily verified by writing the two integration
variables in terms of their real and imaginary parts, i.e. by setting
$z=u+iv$,$z^{\dagger}=u-iv$: the resulting object is a simple Gaussian
integral over the real plane. Thus, Eq. \eqref{eq:integral_transform_complex_generic}
is used to decouple the interaction terms in the partition functions Eq. \eqref{eq:partfunctionBAM}.
introducing a pair of complex conjugate variables for each pattern $\mu$. Note that it is necessary to decouple the extensive noisy part of the patterns, i.e. $\mu\in\left\{ l+1,\ldots,K\right\} $, while
the signal term is left untouched. After applying the transformation \eqref{eq:integral_transform_complex_generic}, 
the partition function \eqref{eq:partfunctionBAM} reads:
\begin{align}
Z\left(\beta,\left\{ \left(\boldsymbol{\xi}^{\mu},\bar{\boldsymbol{\xi}}^{\mu}\right)\right\} _{\mu=1}^{K}\right)=&\sum_{\boldsymbol{\sigma},\bar{\boldsymbol{\sigma}}}\exp\left\{ \frac{\beta}{L}\sum_{\mu=1}^{l}\left(\sum_{i}\xi_{i}^{\mu}\sigma_{i}\right)\left(\sum_{j}\bar{\xi}_{j}^{\mu}\bar{\sigma}_{j}\right)\right\} \times \nonumber \\
&\times\prod_{\mu=l+1}^{K}\int dz_{\mu}z_{\mu}^{\dagger}\exp\left[-\beta z_{\mu}z_{\mu}^{\dagger}+\frac{\beta}{\sqrt{N}}z_{\mu}\sum_{i}\xi_{i}^{\mu}\sigma_{i}+\frac{\beta}{\sqrt{{\N}}}z_{\mu}^{\dagger}\sum_{j}\bar{\xi}_{j}^{\mu}\bar{\sigma}_{j}\right].\label{eq:quenched_p_appendix_after_decouplings}
\end{align}
We must now average $\log Z$ over the quenched disorder to calculate the free energy of the system \eqref{eq:quenched_p_starting}, whose extremization w.r.t. the order parameters will provide a set of self-consistent equations that, in turn, determine the phase diagram of the model. \\
Finally, notice how the integral transform Eq. \eqref{eq:integral_transform_complex_generic} can be exploited to derive a structural analogy between the BAM's partition function and the partition function of two coupled RBMs, by generalizing the well-know equivalence between the Hopfield model and a Binary-Gaussian RBM \citep{barra_equivalence_2012}, further discussed in Appendix \ref{sec:Equivalence_coupledRBMS}.
\subsection{Replica Symmetric phase diagram\label{sec:RS}}
The computation of \eqref{eq:quenched_p_starting}-\eqref{eq:quenched_p_appendix_after_decouplings} is performed using the interpolation technique developed by Guerra~\citep{guerra_course_2006}. This method, restricting its usage to the replica symmetric case analyzed here, allows one to compute the quenched free energy density of a given model whose computation is typically
intractable due to the random interactions in the model's Hamiltonian. The method consists in defining an interpolating free energy function expressed in
terms of an additional parameter $t\in\left[0,1\right]$: when the value of this parameter is $t=0$, the partition function corresponds to a system of independent variables and can be factorized over the spins, which can be computed easily. When $t=1$, we recover the original free energy to be computed (given here by \eqref{eq:quenched_p_starting}). For instance, considering just the signal term in Eqs. \eqref{eq:hamiltonian_separatedpatterns}-\eqref{eq:quenched_p_appendix_after_decouplings}, this is equivalent to consider the following generalized ($t-$dependent) Hamiltonian:
\begin{equation}
 \mathcal{H}^{\rm inter}_t(\boldsymbol{\sigma},\bar{\boldsymbol{\sigma}}) = t\mathcal{H}^{\rm BAM}(\boldsymbol{\sigma},\bar{\boldsymbol{\sigma}})+\left(1-t\right)\beta\sum_{\mu=1}^{K}\left[\Psi_{\mu}\sum_{i}\xi_{i}^{\mu}\sigma_{i}+\bar{\Psi}_{\mu}\sum_{j}\bar{\xi}_{j}^{\mu}\bar{\sigma}_{j}\right].
\end{equation}
 Finally, the method exploits the
fundamental theorem of calculus to compute the free energy at $t=1$
knowing the initial (Cauchy) condition at $t=0$, plus the integral of the derivative of the interpolating free energy with respect to $t$: by assuming replica symmetry, such an integral becomes analytic. It is worth mentioning that this interpolation method has been shown to be rigorous in some disordered models~\citep{barbier_adaptive_2019,Barbier_pedagogic}.
This previous qualitative explanation is expanded and adapted to our model's case in Appendix~\ref{sec:Interpolation-Method}, where all calculation details are explicitly given. The results of this interpolation analysis are described and summarized below.

In the thermodynamic limit, the equilibrium behavior of the BAM is described by a finite number of order parameters, each of which is determined self-consistently by imposing stationary conditions on the resulting free energy.
By analogy with the Hopfield model, we can immediately identify the relevant set of order parameters:
\begin{align}
m_{\mu}=\frac{1}{N}\sum_{i}\xi_{i}^{\mu}\sigma_{i}, & \qquad\qquad\qquad\bar{m}_{\mu}=\frac{1}{{\N}}\sum_{j}\bar{\xi}_{j}^{\mu} \bar{\sigma}_{j};\label{eq:mattis_both}\\
q_{12}=\frac{1}{N}\sum_{i}\sigma_{i}^{\left(1\right)}\sigma_{i}^{\left(2\right)}, & \qquad\qquad\qquad\bar{q}_{12}=\frac{1}{{\N}}\sum_{j}\bar{\sigma}_{j}^{\left(1\right)}\bar{\sigma}_{j}^{\left(2\right)};\label{eq:overlapdef_both}\\
p_{12}=\frac{1}{K}\sum_{\mu>l}^{K}z_{\mu}^{\left(1\right)}z_{\mu}^{\left(2\right)}, & \qquad\qquad\qquad\bar{p}_{12}=\frac{1}{K}\sum_{\mu>l}^{K}z_{\mu}^{\dagger\left(1\right)}z_{\mu}^{\dagger\left(2\right)}.\label{eq:overlapP_def_both}
\end{align}
The first line reminds the set of \textit{Mattis magnetizations}, representing the projection of a spin configuration (on either one of the two layers) along one of the retrievable patterns, i.e. $\mu\in\left\{ 1,\ldots,l\right\}$.  By using
the short-hand notation $\boldsymbol{m}=\left(m_{1},\ldots,m_{l}\right)^{T}$
and $\bar{\boldsymbol{m}}=\left(\bar{m}_{1},\ldots,\bar{m}_{l}\right)^{T}$,
in the retrieval phase, at least one component of each of these vectors (in particular, those with the same index) is expected to be nonzero, so that the free
energy minima are characterized by spin configurations with a finite
overlap with a particular pattern pair. The second line defines the standard
overlap matrix between two replicas (here indexed by $1$, $2$)
for each layer. Finally, the last line defines a set of overlaps between the complex variables introduced by the the integral transform~\eqref{eq:quenched_p_appendix_after_decouplings}. Under the replica symmetric (RS) assumption, each of these order parameters does not fluctuate in the thermodynamic limit and its distribution reaches a delta-function peaked around a unique value determined by imposing stationarity on the free energy, namely
\begin{equation}
\lim_{L\to \infty} P\left( \left \langle \mathcal{O} \right \rangle \right) = \delta \left(\left \langle \mathcal{O} \right \rangle - O \right)\label{eq:poverlapRS} 
\end{equation}
where $\mathcal{O}$ denotes any of the order parameters defined in Eqs. \eqref{eq:mattis_both}-\eqref{eq:overlapdef_both}-\eqref{eq:overlapP_def_both}, and $O$ denotes the corresponding equilibrium value.
To simplify the notation, in the following we highlight the RS value of each of the above order parameters with a capital letter, to be distinguished from the general expression in small.  
The RS free energy density of the BAM is given by the following expression (see Appendix \ref{sec:Interpolation-Method} for calculation details):
\begin{align}
f\left(\beta,\alpha,\gamma\right) =&\;  \sum_{\mu=1}^{l}M_{\mu}\bar{M}_{\mu}+\frac{\alpha\beta}{2}P\left(1-Q\right)+\frac{\alpha\beta}{2}\bar{P}\left(1-\bar{Q}\right)+\nonumber \\
 & -\frac{\gamma}{\beta}\mathbb{E}_{\eta,\boldsymbol{\xi}}\log2\text{cosh}\left[\beta\sqrt{\bar{\gamma}\alpha P}\eta+\beta\bar{\gamma}\sum_{\mu}\bar{M}_{\mu}\xi^{\mu}\right]-\frac{\bar{\gamma}}{\beta}\mathbb{E}_{\bar{\eta},\bar{\boldsymbol{\xi}}}\log2\text{cosh}\left[\beta\sqrt{\gamma\alpha\bar{P}}\bar{\eta}+\beta\gamma\sum_{\mu}M_{\mu}\bar{\xi}^{\mu}\right]+\nonumber \\
 & +\frac{\alpha}{2\beta}\log\Delta-\frac{\alpha\beta}{2\Delta}\left[Q\left(1-\bar{Q}\right)+\bar{Q}\left(1-Q\right)\right],\label{eq:RS_freeenergy}
\end{align}
where $\Delta=1-\beta^{2}\left(1-Q\right)\left(1-\bar{Q}\right)$, 
$\bar{\gamma}=\gamma^{-1}$ and $\eta,\bar{\eta}$ are two i.i.d. standard Gaussian variables.
The above expression requires $\Delta$ to be always positive: such a condition is always verified at any point of the phase diagram. 
We stress that the same result can be obtained using the more standard saddle point method through the replica trick \citep{mezard1987spin}, as originally carried out in \citep{kurchan_statistical_1994}. However, in Ref. \citep{kurchan_statistical_1994} there clearly was a misprint or a small error in earlier calculations (for instance, the sign of $\Delta$ was exchanged, leading to a free-energy being ill-defined at finite $\beta$) as results were not consistent with our numerical simulations (discussed in Section \ref{sec:numericalsim}).

The values of the order parameters at fixed control parameters can be obtained by evaluating the saddle points of $f$
of Eq. \eqref{eq:RS_freeenergy}. 
The set of self-consistent equations to be fulfilled by these values is shown below:
\begin{subequations}
\begin{alignat}{2}
\boldsymbol{M} & = \mathbb{E}_{\eta,\boldsymbol{\xi}}\; \boldsymbol{\xi}\text{tanh}\left[\beta\sqrt{\bar{\gamma}\alpha P}\eta+\beta\bar{\gamma}\left(\boldsymbol{\xi}\cdot\bar{\boldsymbol{M}}\right)\right]\, ,\qquad\qquad\qquad &\bar{\boldsymbol{M}}&= \mathbb{E}_{\bar{\eta},\bar{\boldsymbol{\xi}}} \; \bar{\boldsymbol{\xi}} \text{tanh}\left[\beta\sqrt{\gamma\alpha\bar{P}}\bar{\eta}+\beta\gamma\left(\bar{\boldsymbol{\xi}}\cdot\boldsymbol{M}\right)\right];\label{eq:SP_Mbvec_rhs}\\
Q & =\mathbb{E}_{\eta,\boldsymbol{\xi}} \; \text{tanh}^{2}\left[\beta\sqrt{\bar{\gamma}\alpha P}\eta+\beta\bar{\gamma}\left(\boldsymbol{\xi}\cdot\bar{\boldsymbol{M}}\right)\right]\, , & \bar{Q}&=\mathbb{E}_{\bar{\eta},\bar{\boldsymbol{\xi}}}\; \text{tanh}^{2}\left[\beta\sqrt{\gamma\alpha\bar{P}}\eta+\beta\gamma\left(\bar{\boldsymbol{\xi}}\cdot\boldsymbol{M}\right)\right];\label{eq:SP_Qb_rhs}\\
P & =\Delta^{-2}\left[\bar{Q}+\beta^{2}Q\left(1-\bar{Q}\right)^{2}\right]\, , & \bar{P} &=\Delta^{-2}\left[Q+\beta^{2}\bar{Q}\left(1-Q\right)^{2}\right].\label{eq:SP_Pb_rhs}
\end{alignat}
\end{subequations}
\begin{figure}
\begin{overpic}[width=\textwidth]{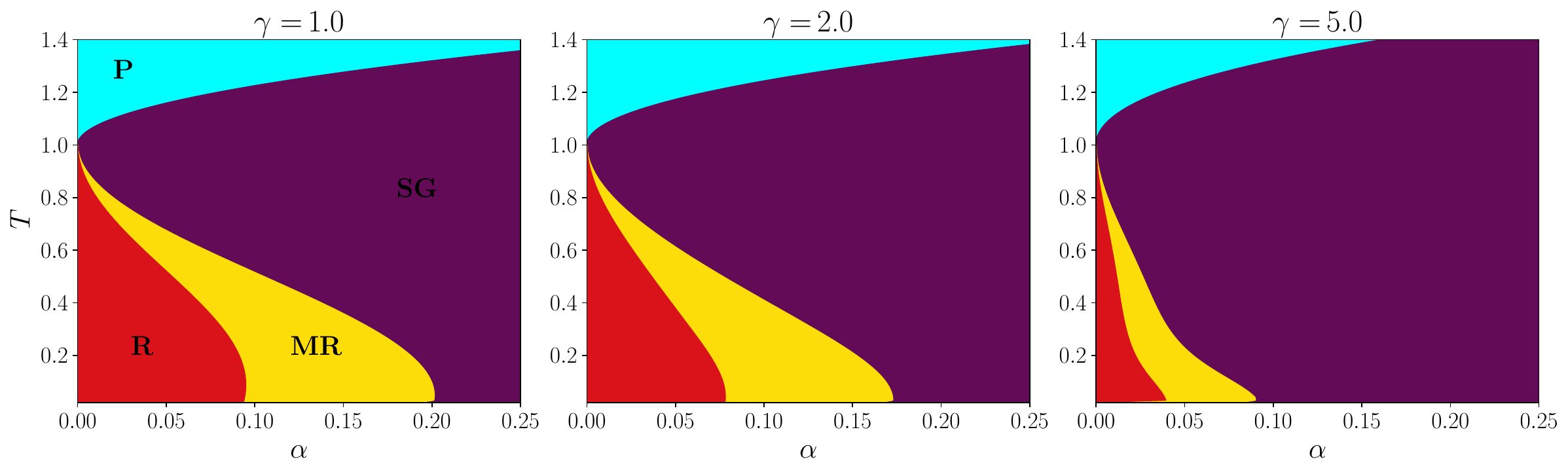}
\put(1,30){{\figpanel{a}}}
\put(35,30){{\figpanel{b}}}
\put(69,30){{\figpanel{c}}}
\end{overpic}
\caption{Phase diagram of the BAM in the plane $\left(\alpha,T\right)$ at
different values of $\gamma$. \figpanel{a}: $\gamma=1$. \figpanel{b}: $\gamma=2$; \figpanel{c}: $\gamma=5$. The labels $\textbf{P}$, $\textbf{SG}$,
$\textbf{R}$, $\textbf{MR}$ (shown only in \figpanel{a}) stand for paramagnetic, spin-glass,
retrieval and metastable-retrieval, respectively. \label{fig:Phasediagrams_all}}
\end{figure}
The model phase diagram can be fully characterized by solving the
above saddle point equations \eqref{eq:SP_Mbvec_rhs}-\eqref{eq:SP_Pb_rhs} at any value of the
three control parameters $\left(\alpha,\beta,\gamma\right)$, and evaluating the corresponding free energy for each of the fixed points.  For
simplicity, we just consider the retrieval of $1$ pattern pair: this is equivalent to assume that only one component of the
vectors $\boldsymbol{M},\bar{\boldsymbol{M}}$ is non-zero, namely
$M_\nu =M_{\mu}\delta_{\mu,\nu}$ and $\bar{M}_\nu = \bar{M}_{\mu}\delta_{\mu,\nu}$ with $\mu$ being the pattern index to be retrieved (one could set
$\mu=1$ without loss of generality).

We start from the symmetric case where $\gamma=1$ (we recall the parameter $\gamma$ controls the ratio between the size of the two layers, see \eqref{eq:gamma}), and the corresponding phase diagram is shown in Fig. \ref{fig:Phasediagrams_all} \figpanel{a} in the plane $\left( \alpha, T=\beta^{-1}\right)$. 

At high temperature, the equilibrium phase is \textit{paramagnetic} ($\textsc{P}$) with all the order parameters being null. At $T<1$ and sufficiently small loads $\alpha$, the model exhibits a ferromagnetic-like, so-called \textit{retrieval} ($\textsc{R}$)
phase, with all the order parameters being non-null. Further increasing
the load $\alpha$ over a certain threshold $\alpha_{c}\left(T\right)$
marks the onset of a \textit{spin-glass} (SG) phase with zero Mattis magnetizations
over all patterns but non-zero overlaps $Q,\bar{Q}\neq0$. A more detailed inspection shows how the retrieval phase can be divided in two sub-regimes:
a first (red) portion of the phase diagram where the retrieval fixed
point has the lowest free energy, and another one (yellow) where the
retrieval fixed point is a local stationary point. In the former, while the retrieval phase is dominant, the spin-glass regime is still locally stable. In the latter,
the global minimum of the free energy is the spin-glass fixed point: however, the retrieval solution is locally stable (metastable, MR), so that it is still possible to find it by starting with an initial condition (either in the saddle point equations or in the neural dynamics for finite size systems) that has a sufficiently large overlap with a pattern pair in the retrievable set.

In the asymmetric case ($\gamma\neq1$) the phase diagram looks qualitatively
similar to the previous one, as clear from Fig. \ref{fig:Phasediagrams_all} \figpanel{b}-\figpanel{c}, showing respectively the result
at $\gamma=2$ and $\gamma=5$. However, the retrieval region of the
diagram shrinks as $\gamma$ departs from $1$ (i.e. from the completely symmetric case), so that retrieval
loss occurs at lower values of $\alpha$. Moreover, the critical line
separating the paramagnetic from the spin-glass phases moves to higher temperatures. As an overall remark, we can conclude that increasing the asymmetry between the two layers extends the spin-glass phase at the expense of the others. An important property to notice is that the above free energy (and consequently, the phase diagram) is symmetric
under the exchange of the two layers, provided that all the order parameters are swapped and $\gamma\leftrightarrow\bar{\gamma} = 1 \slash \gamma $.

\begin{figure}
\begin{overpic}[width=\textwidth]{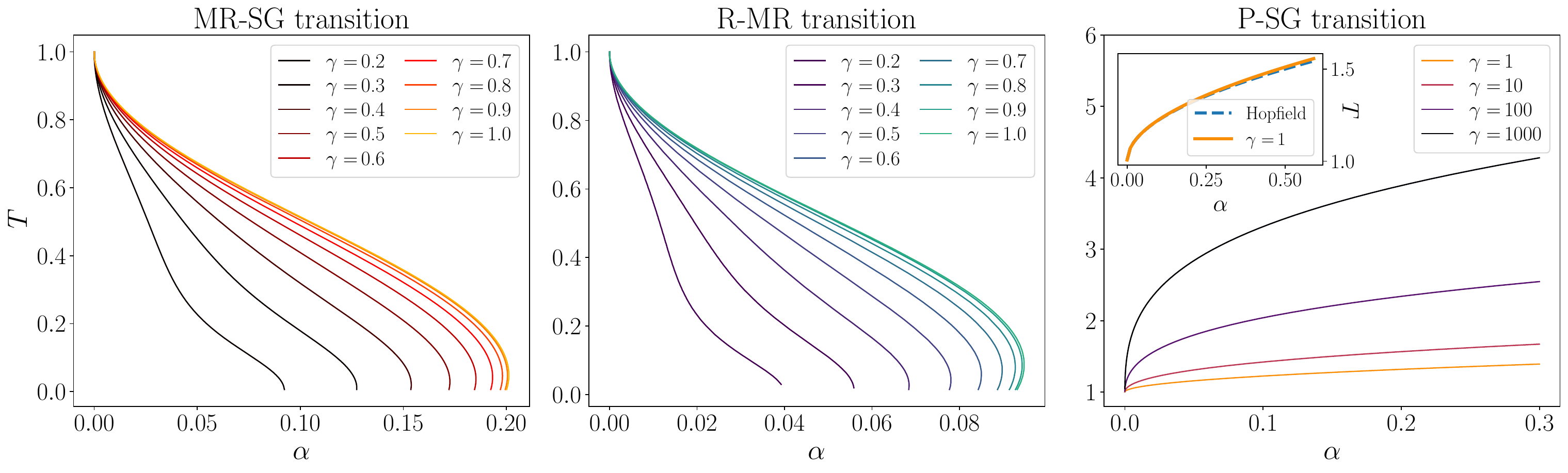}
\put(2.5,30){{\figpanel{a}}}
\put(35,30){{\figpanel{b}}}
\put(68.5,30){{\figpanel{c}}}
\end{overpic}

\caption{Critical lines separating the different operating regimes of the BAM. In \figpanel{a}, critical lines for the MR-SG. In \figpanel{b}, critical lines for the R-MR transition at different values of $\gamma$.
In \figpanel{c}, critical lines for the P-SG transition at different values of $\gamma$. All the lines are plotted as function of $\alpha$ (please mind the three panels cover different ranges of $\alpha$), for different values of $\gamma$.
The inset in panel \figpanel{c} shows a comparison of the P-SG critical lines between the symmetric BAM, given by \eqref{eq:T_SG-P_line_symmetric}, and the same critical line computed for the Hopfield model (after a proper rescaling to take into account the different definition of the load in the two models). \label{fig:allcritical_lines}}
\end{figure}

A numerical characterization of the transition lines for different values of $\gamma$ is shown in Figure \ref{fig:allcritical_lines}: the metastable retrieval-spin glass (MR-SG) in \figpanel{a}, the stable-metastable retrieval (R-MR) in \figpanel{b}, and the paramagnetic-spin glass (P-SG) in \figpanel{c}. In all cases, we find that the greater the asymmetry of the BAM, the narrower the retrieval region and the wider the spin glass.
These three transition lines are obtained as follows: the first-order spinodal of the retrieval state, corresponding to the line between the  metastable-retrieval phase and the spin glass one, is found by solving the saddle point equations starting from a large overlap with one pattern, until such a solution ceases to exist upon increasing $\alpha$. The critical line between the stable and metastable retrieval phase is instead found by looking at the points where the free energies of the retrieval state and the spin glass solution (the latter obtained by setting $M=\bar{M}=0$ in the saddle point equations) are equal. Finally, the P-SG line allows an analytical description and will therefore discussed in detail in the next paragraph.
We finally notice that in Fig. \ref{fig:allcritical_lines} \figpanel{a}-\figpanel{b} the transition lines' reentrance at small values of $T$ is due to an instability of the retrieval solution in the replica space, evidencing the failure of the RS ansatz in this regime, analogously to the Hopfield model. This was noted in \citep{tanaka2000capacity} in the symmetric case $\gamma = 1$. For completeness, in the next section we investigate the low-temperature behavior of the BAM using a 1-RSB ansatz.

\paragraph{P-SG transition line}
The critical line between the paramagnetic and spin-glass phases can be further analyzed analytically because it is of second order. We
proceed by setting $\boldsymbol{M}=\bar{\boldsymbol{M}}=\boldsymbol{0}$
into the saddle-point equations and expand Eqs. \eqref{eq:SP_Qb_rhs}
around $Q,\bar{Q}\approx0$. 
To the first order, the Jacobian of the linearized systems and its eigenvalues are shown below:
\begin{equation}
    \mathcal{J}  =\frac{\alpha\beta^{2}}{\left(1-\beta^{2}\right)^{2}}\begin{bmatrix}\bar{\gamma}\beta^{2} & \bar{\gamma}\\
\gamma & \gamma\beta^{2}
\end{bmatrix} \quad \Longrightarrow \quad \lambda_{\pm}\left(\alpha,\beta,\gamma\right)=\frac{\alpha\beta^{2}}{2\gamma\left(1-\beta^{2}\right)^{2}}\left[\beta^{2}\left(1+\gamma^{2}\right)\pm\sqrt{4\gamma^{2}+\beta^{4}\left(1-\gamma^{2}\right)^{2}}.\right]\label{eq:eigenvalues_Jac_SG-Pline}
\end{equation}
At fixed $\gamma$, the critical line in the plane $\left(\alpha,T\right)$
is given by the first eigenvalue becoming larger than $1$ upon lowering
the temperature. In particular, in the symmetric case $\gamma=1$,
the equation $\lambda_{\pm}\left(\alpha,\beta,\gamma=1\right)=1$
can be solved analytically, leading to the following expression for
the critical temperature:
\begin{align}
T_{\text{P-SG}}\left(\alpha,\gamma=1\right) & =\sqrt{1+\frac{\alpha}{2}+\frac{1}{2}\sqrt{\alpha\left(\alpha+8\right)}}.\label{eq:T_SG-P_line_symmetric}
\end{align}
A comparison between \eqref{eq:T_SG-P_line_symmetric} and the Hopfield
critical line as predicted by the AGS theory \citep{amit_storing_1985}
is shown in the inset of Fig. \ref{fig:allcritical_lines} \figpanel{c}: however, attention must be paid to the different definitions of $\alpha$ in the two models. In the Hopfield case, the network load needs to be defined as $\alpha^{\text{Hopfield}} = K / (N + \N )$ in order to be consistent with the overall dimension of patterns in both models. In the symmetric case where $\gamma=1$, this implies $\alpha^{\text{Hopfield}} = \alpha^{\text{BAM}} \slash 2$.
As a consequence, the Hopfield's critical line plotted in Fig. \ref{fig:allcritical_lines} corresponds to $T_c^{\text{Hopfield}} = 1+ \sqrt{\alpha^{\text{BAM}} \slash 2}$, showing a very similar behaviour between the two curves (they only depart at very high values of $\alpha$).
In the asymmetric case, critical lines can be found numerically by solving $\lambda_{\pm}\left(\alpha,\beta,\gamma\right)=1$
at different values of $\gamma$ and $\alpha$, plotted in Fig.~\ref{fig:allcritical_lines} \figpanel{c}. Notice how, at fixed load $\alpha$, such a phase transition is pushed at higher temperature when the asymmetry between the two layers increases.

\subsubsection{RS phase diagram at $T=0$\label{subsec:RSphasediagramT0}}
\begin{figure}
\centering
\begin{overpic}[width=.9\textwidth]{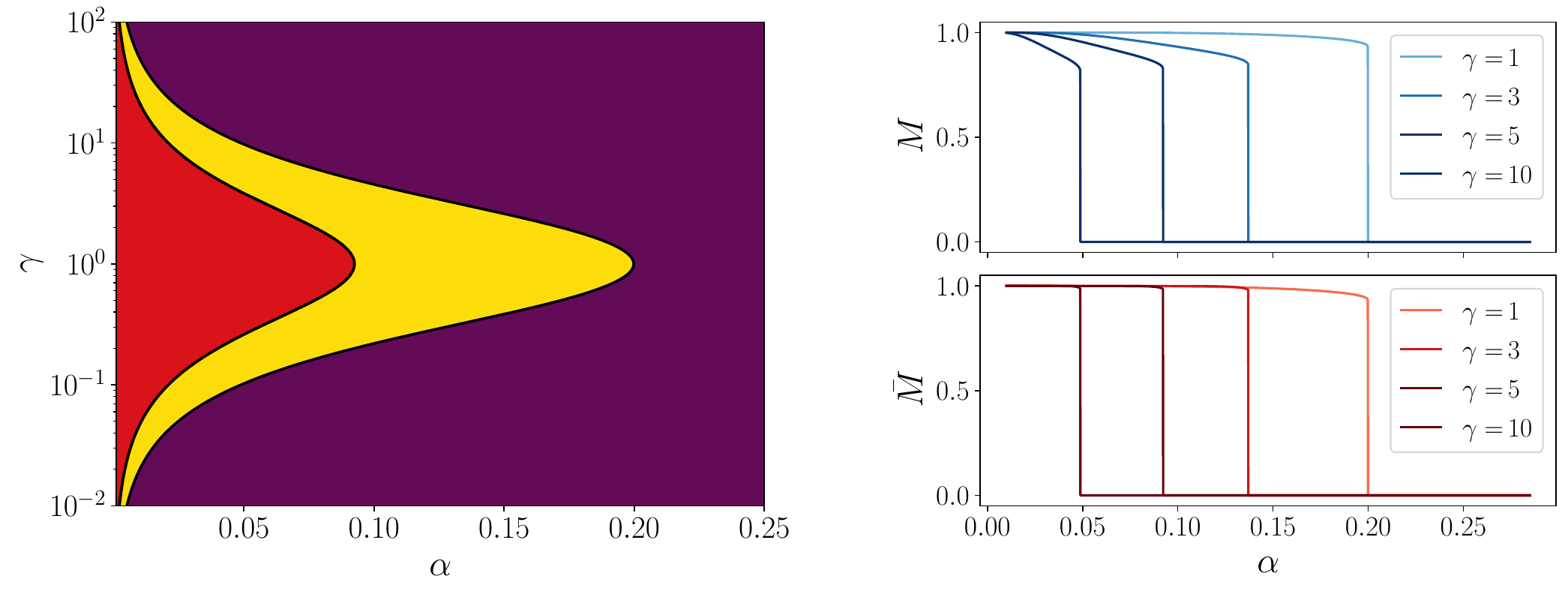}
\put(0.5,35){{\figpanel{a}}}
\put(56,35){{\figpanel{b}}}
\put(56,19){{\figpanel{c}}}
\end{overpic}

\caption{Panel \figpanel{a}: BAM's phase diagram at $T=0$ in the plane $\left(\alpha,\gamma\right)$: the same color-code as in Figure \ref{fig:Phasediagrams_all} is used. The phase boundary separating the MR (yellow) from the SG (violet) phases defines the RS critical capacity $\alpha_{c}$
as a function of $\gamma$. As the vertical axis has a logarithmic scale,
we notice again how the phase diagram is symmetric under the transformation
$\gamma\to\gamma^{-1}=\bar{\gamma}$. Panels \figpanel{b} and \figpanel{c} show the zero-temperature Mattis magnetizations of the two layers (plotted resp. in panels \figpanel{b} and \figpanel{c}) as functions of $\alpha$ for $4$ different values of $\gamma$. \label{fig:phasediagram_T0}}
\end{figure}

\begin{figure}
\centering
\begin{overpic}[width=.9\textwidth]{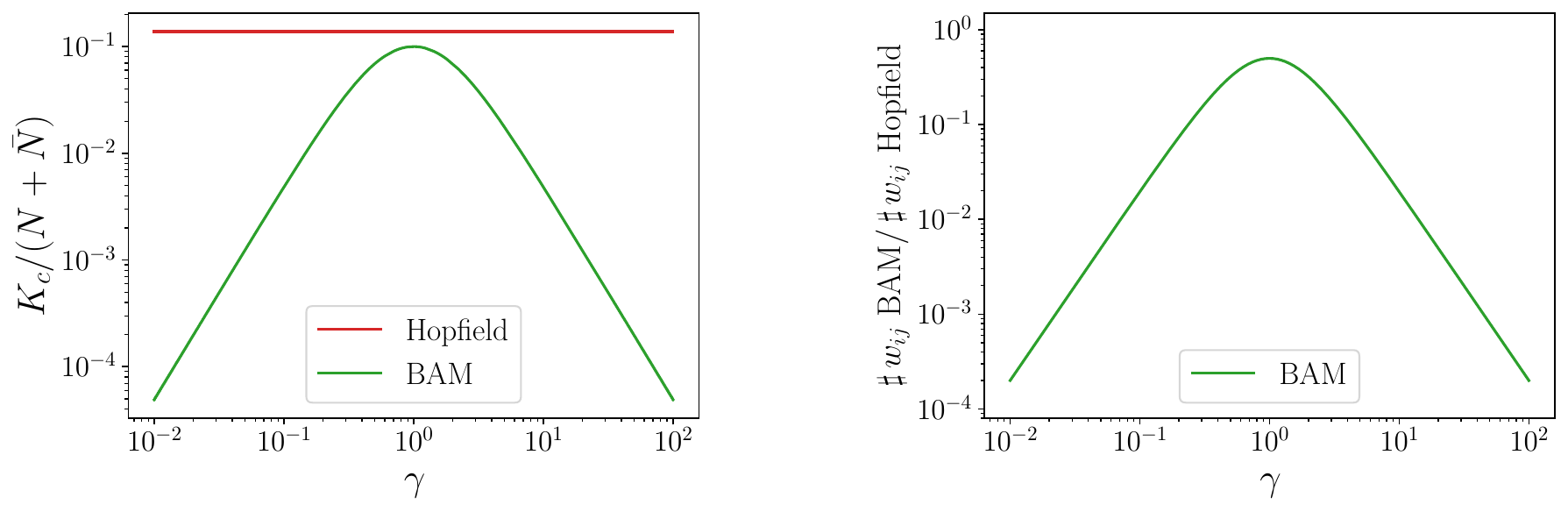}
\put(0.5,32){{\figpanel{a}}}
\put(55,32){{\figpanel{b}}}
\end{overpic}
\caption{\figpanel{a} Critical capacity  of BAM and the Hopfield but normalized by the total number of neurons in the network (i.e. $N+\N$ instead of $L$ as in \eqref{eq:alpha} for the BAM). The green line corresponds to Eq. \eqref{eq:criticalcapBAM_renormalized}.
\figpanel{b} Number of weights used to store this critical number of patterns in the BAM normalized by the analogous number of weights in the Hopfield model\footnote{Note that this quotient is equal to $N \N \slash \binom{N+\N}{2}$.}. In both panels, each quantity is plotted as a function of the asymmetry $\gamma = \sqrt{N\slash \N}$. \label{fig:comparisonBAMHopfield}}
\end{figure}

The phase diagram in the noiseless regime can be computed by taking
the $T=0$ limit of the free energy \eqref{eq:RS_freeenergy} and/or
the corresponding saddle point equations (\ref{eq:SP_Mbvec_rhs}-\ref{eq:SP_Pb_rhs}). The
calculations follow the same reasoning as discussed in \citep{kurchan_statistical_1994},
here we just report the final result. Again, we focus on the retrieval of one pattern pair, so that the vectors $\boldsymbol{M}$ and
$\bar{\boldsymbol{M}}$ defined in Eq \eqref{eq:mattis_both} have
only one nonzero component by construction. At $\beta\to\infty$, the overlaps
$Q,\bar{Q}$ tend to $1$, in such a way that the spin-glass susceptibilities
on the two spin sets $\chi=\beta\left(1-Q\right)$ and $\bar{\chi}=\beta\left(1-\bar{Q}\right)$
maintain finite values. The saddle-point equations can therefore be
re-arranged in a different form, to determine self-consistently the
susceptibilities and the Mattis magnetizations of each layer. After
taking the zero-temperature limit, the new saddle-point equations
read:
\begin{align}
\frac{1+\bar{\chi}^{2}}{\left(1-\chi\bar{\chi}\right)^{2}} & =\frac{\text{erf}^{2}\left(\bar{y}\right)}{2\gamma\alpha y^{2}},\qquad\qquad\qquad\chi=\frac{2}{\bar{\gamma}\sqrt{\pi}}\frac{y}{\text{erf}\left(\bar{y}\right)}e^{-y^{2}},\label{eq:sp_t0_1-1}\\
\frac{1+\chi^{2}}{\left(1-\chi\bar{\chi}\right)^{2}} & =\frac{\text{erf}^{2}\left(y\right)}{2\bar{\gamma}\alpha\bar{y}^{2}},\qquad\qquad\qquad\bar{\chi}=\frac{2}{\gamma\sqrt{\pi}}\frac{\bar{y}}{\text{erf}\left(y\right)}e^{-\bar{y}^{2}},\label{eq:sp_t0_2-1}
\end{align}
where $\text{erf}\left(\cdot \right)$ denotes the error function, and the variables $y,\bar{y}$ are linked to the Mattis magnetizations of each layer through
\begin{equation}
M=\text{erf\ensuremath{\left(y\right)}},\qquad\qquad\qquad\bar{M}=\text{erf}\left(\bar{y}\right).\label{eq:mattis0_both}
\end{equation}
The above system can be solved w.r.t. the variables $y,\bar{y}$ at
different values of $\alpha,\gamma$. The resulting phase diagram
is shown in Figure \ref{fig:phasediagram_T0} \figpanel{a}. The gray line separating
the MR phase (in yellow) and the spin-glass phase (violet)
defines the critical storage capacity $\alpha_{c}\left(\gamma\right)$
of the BAM. Notice how the maximum value is reached at $\gamma=1$,
where $\alpha_{c}\left(\gamma=1\right)\approx0.2$, consistently with  \citep{englisch_bm_1995,tanaka2000capacity}. Intuitively, this is understood by considering the number of weights encoded in the synaptic matrix \eqref{eq:coupling_matrix} of the BAM, which is equal to $N \N = L^2$. Considering a network with a fixed number of neurons $N+\N$ (or equivalently, with a fixed pattern dimension), it can be immediately verified that the maximum number of weight entries in the BAM is reached when $\gamma = 1$. The general idea is that the more weights there are in the network, the larger the retrieval phase with respect to the quenched noise induced by the extensive number of patterns stored in the network. We also mention how in the extremely asymmetric limit where e.g. $\gamma\to0$ the storage capacity $\alpha_c$ goes to $0$ linearly with $\gamma$, so that the rescaled capacity $\tilde{\alpha}_c = \alpha_c \slash \gamma$ tends to a finite value $\tilde{\alpha}_c \to 0.497$. By symmetry, the same result is achieved when $\gamma\to \infty$, and defining the rescaled capacity as $\tilde{\alpha}_c = \gamma\alpha_c = \alpha_c \slash \bar{\gamma}$. 
Figures \ref{fig:phasediagram_T0} \figpanel{b}-\figpanel{c} instead show $4$ horizontal cuts of the phase diagram at different values of $\gamma$ (as indicated in the caption). Specifically, in each panel we plot the Mattis magnetization of the fixed point for the two layers (in \figpanel{b}-\figpanel{c} respectively) as a function of $\alpha$, showing a first-order transition between the retrieval and spin-glass phase. Interestingly, the Mattis magnetization of the largest layer, show in Fig. \ref{fig:phasediagram_T0} \figpanel{b}, decreases continuously before the critical value $\alpha_c$.\\
We finally quantify the difference between the Hopfield model and the BAM in terms of their storage capabilities: in order to have a fair comparison between the two models, it is more convenient to renormalize their critical capacities by the total number of neurons $N+\bar{N}$, which is kept fixed whilst varying the asymmetry between the two layers. In this notation, while the Hopfield's critical capacity is equal to $\alpha_c^{\text{Hopfield}} = K_c \slash (N+\bar N) \approx 0.138 $ independenty on $\gamma$, the BAM's critical capacity is given by
\begin{equation}
\tilde{\alpha}_c^{\text{BAM}} = \frac{K_c^{\text{BAM}}}{N+\N} =  \frac{\alpha_c \left(\gamma \right)}{\gamma + \gamma^{-1}},\label{eq:criticalcapBAM_renormalized}
\end{equation}
where at the right-most hand side $\alpha_c$ is the quantity plotted in Figure \ref{fig:phasediagram_T0} \figpanel{a}. These two quantities are shown in Figure \ref{fig:comparisonBAMHopfield} \figpanel{a}: as expected, we conclude that BAM's critical capacity is lower than the Hopfield model even in its most stable configuration (i.e. the symmetric case). 
On the other hand, the BAM is more efficient from the point of view of the number of weights stored in the network, since for a fixed number of patterns such that the network operates in its retrieval phase, it requires (at most, in the symmetric case) about half of the weights compared to the Hopfield model (see Figure \ref{fig:comparisonBAMHopfield} \figpanel{b}).

\subsection{1-RSB analysis\label{sec:1RSB}}
A more peculiar investigation of the phase diagram requires analyzing the stability of the RS solution. For the Hopfield model, while the spin-glass solution of the RS fixed point equations is always be unstable below the P-SG critical line (below which the system is not anymore ergodic), the instability of the retrieval state occurs only at very low temperature close to the spinodal curve of the MR-SG \citep{amit_storing_1985,coolen_statistical_2000}, below the so-called de Almeida Thouless (dAT) line \citep{dATSK}. For the BAM, this phenomenon was already observed \citep{tanaka2000capacity} in the symmetric scenario where $\gamma=1$ (generalization to the asymmetric case being straightforward).
Therefore, we now perform a 1-step replica symmetry breaking (1-RSB) analysis of the BAM to investigate how it impacts the critical capacity of the model. In general, this ansatz implies that the overlap distribution acquires a bi-modal structure with two equilibrium values, and this property holds for any of the four overlaps defined in Eqs. \eqref{eq:overlapdef_both}-\eqref{eq:overlapP_def_both}. Specifically, let us denote with $\mathcal{O}$ either one of the four overlaps $q_{12}, \bar{q}_{12}, p_{12}, \bar{p}_{12}$, its probability distribution will be given in the thermodynamic limit by
\begin{equation}
    \lim_{L\to \infty} P \left (\left \langle \mathcal{O} \right\rangle \right) = \theta \delta \left( \left\langle\mathcal{O} \right\rangle- O_1 \right) + \left(1-\theta\right) \delta \left( \left\langle\mathcal{O} \right\rangle- O_2\right)
\end{equation}
where $O_{1},O_{2}$ are the two equilibrium values weighted by $\theta \in \left[0,1\right]$. On the other hand, magnetizations still are supposed to self-average as in the RS case (c.f. Eq \eqref{eq:poverlapRS})
The computation of the 1-RSB free energy can be easily carried out either by using the replica trick at the 1-RSB level or by extending the technique presented in Ref. \citep{Agliari_rsb} where an interpolation method at the 1-RSB level is presented for the Hopfield model. Regarding the BAM, one just needs to adapt the definition of the interpolating pressure as discussed in Appendix \ref{sec:Interpolation-Method} for the RS case. In either cases, and assuming and $O_2 >O_1$ for any of the four overlaps in Eqs. \eqref{eq:overlapdef_both}-\eqref{eq:overlapP_def_both} the final result reads
\begin{align}
f\left(\beta,\alpha,\gamma\right) \, = & \sum_{\mu=1}^{l}M_{\mu}\bar{M}_{\mu}+\frac{\alpha\beta}{2}P_{2}\left[1+\left(\theta-1\right)Q_{2}\right]+\frac{\alpha\beta}{2}\bar{P}_{2}\left[1+\left(\theta-1\right)\bar{Q}_{2}\right]-\frac{\alpha\beta}{2}\theta\left[Q_{1}P_{1}+\bar{Q}_{1}\bar{P}_{1}\right] + \nonumber\\
 & +\frac{\alpha}{2\beta}\log\Delta+\frac{\alpha}{2\beta\theta}\log\frac{\Delta_{\theta}}{\Delta}-\frac{\alpha}{2\Delta_{\theta}}\left(\bar{Q}_{1}\kappa+Q_{1}\bar{\kappa}\right)-\frac{\gamma}{\beta\theta}\mathbb{E}_{\boldsymbol{\xi},\tau}\log\mathbb{E}_{z}\text{2}^{\theta}\text{cosh}^{\theta}\Xi-\frac{\bar{\gamma}}{\beta\theta}\mathbb{E}_{\bar{\boldsymbol{\xi}},\tau}\log\mathbb{E}_{z}\text{2}^{\theta}\text{cosh}^{\theta}\bar{\Xi} \label{eq:1RSBfreeen}
\end{align}
where we defined
\begin{equation}
\Xi =\beta\left[\bar{\gamma}\left(\bar{\boldsymbol{M}}\cdot\boldsymbol{\xi}\right)+\sqrt{\bar{\gamma}\alpha P_{1}}\tau+\sqrt{\bar{\gamma}\alpha\left(P_{2}-P_{1}\right)}z\right], \qquad \qquad \bar{\Xi} =\beta\left[\gamma\left(\boldsymbol{M}\cdot\bar{\boldsymbol{\xi}}\right)+\sqrt{\gamma\alpha\bar{P}_{1}}\tau+\sqrt{\gamma\alpha\left(\bar{P}_{2}-\bar{P}_{1}\right)}z\right],\label{eq:XiandXibar}
\end{equation}
and
\begin{subequations}
\begin{align}
\kappa & =\beta\left(1-Q_{2}\right)+\beta\theta\delta Q\\
\bar{\kappa} & =\beta\left(1-\bar{Q}_{2}\right)+\beta\theta \delta\bar{Q}\\
\Delta_{\theta} & =1-\kappa\bar{\kappa},\label{eq:deltatheta1RSB}
\end{align}\label{eq:Kappas1RSB}
\end{subequations}
where $\delta Q = Q_2 - Q_1$, $\delta\bar{Q}  = \bar{Q}_2 - \bar{Q}_1$. 
Finally, $\Delta$ has the same expression as in the previous section (see Eq. \eqref{eq:RS_freeenergy}).
Note that the RS free-energy is recovered in the limit $O_2-O_1\to 0$ with $O \in \{Q,\bar{Q}, P, \bar{P}\}$, or alternatively in one of the two extreme values of $\theta = 0,1$ where the dependency on one of the two values assumed by each overlap disappears.
Once again, the free energy solely depends on the three control parameters $\left(\alpha, \beta, \gamma \right)$, the value of all the order parameters being determined self-consistently through the following equations:
\begin{subequations}
\begin{alignat}{2}
\boldsymbol{M} & = \mathbb{E}_{\boldsymbol{\xi},\tau} \,\boldsymbol{\xi}\, \frac{\mathbb{E}_{z}\,\text{tanh}\, \Xi\,\text{cosh}^{\theta} \Xi}{\mathbb{E}_{z}\,\text{cosh}^{\theta}\Xi}\, ,\qquad\qquad\qquad\quad \qquad \qquad & \boldsymbol{M} & = \mathbb{E}_{\bar{\boldsymbol{\xi}},\tau} \,\bar{\boldsymbol{\xi}} \,\frac{\mathbb{E}_{z}\,\text{tanh}\bar{\Xi}\,\text{cosh}^{\theta}\bar{\Xi}}{\mathbb{E}_{z}\,\text{cosh}^{\theta}\bar{\Xi}}\, ,\\
Q_{1} & =\mathbb{E}_{\boldsymbol{\xi},\tau}\left[\frac{\mathbb{E}_{z}\, \text{tanh}\, \Xi\, \text{cosh}^{\theta}\Xi}{\mathbb{E}_{z}\, \text{cosh}^{\theta}\Xi}\right]^{2}\, , & \bar{Q}_{1} & =\mathbb{E}_{\bar{\boldsymbol{\xi}},\tau}\left[\frac{\mathbb{E}_{z}\, \text{tanh}\bar{\Xi} \, \text{cosh}^{\theta}\bar{\Xi}}{\mathbb{E}_{z}\, \text{cosh}^{\theta}\bar{\Xi}}\right]^{2}\, ,\\
Q_{2} & =\mathbb{E}_{\boldsymbol{\xi},\tau}\, \frac{\mathbb{E}_{z}\, \text{tanh}^{2}\Xi\text{cosh}^{\theta}\Xi}{\mathbb{E}_{z}\, \text{cosh}^{\theta}\Xi}\, , & \bar{Q}_{2} & =\mathbb{E}_{\bar{\boldsymbol{\xi}},\tau}\frac{\mathbb{E}_{z}\, \text{tanh}^{2}\bar{\Xi}\, \text{cosh}^{\theta}\bar{\Xi}}{\mathbb{E}_{z}\, \text{cosh}^{\theta}\bar{\Xi}}\, ,\\
P_{1} & =\Delta_{\theta}^{-2} \left(\bar{Q}_{1}+\bar{\kappa}^{2}Q_{1}\right)\, , & \bar{P}_{1} & =\Delta_{\theta}^{-2} \left(Q_{1}+\kappa^{2}\bar{Q}_{1} \right)\, ,\\
P_{2} & =P_{1}+\left(\Delta \Delta_{\theta}\right)^{-1}\left[\delta\bar{Q}+\beta\left(1-\bar{Q}_{2}\right)\bar{\kappa}\delta Q\right]\, , &  \bar{P}_{2} & =\bar{P}_{1}+\left(\Delta \Delta_{\theta}\right)^{-1}\left[\delta Q+\beta\left(1-Q_{2}\right)\kappa \delta\bar{Q}\right]\, .
\end{alignat}\label{eq:selfconsistent1RSB}
\end{subequations}
Notice that also $\theta$ plays the role of an order parameter, whose equilibrium value is determined by another stationary condition obtained by taking $\partial f \slash \partial \theta = 0$.
\subsubsection{1-RSB phase diagram at $T=0$}
\begin{figure}
\centering
\begin{overpic}[width=\textwidth]{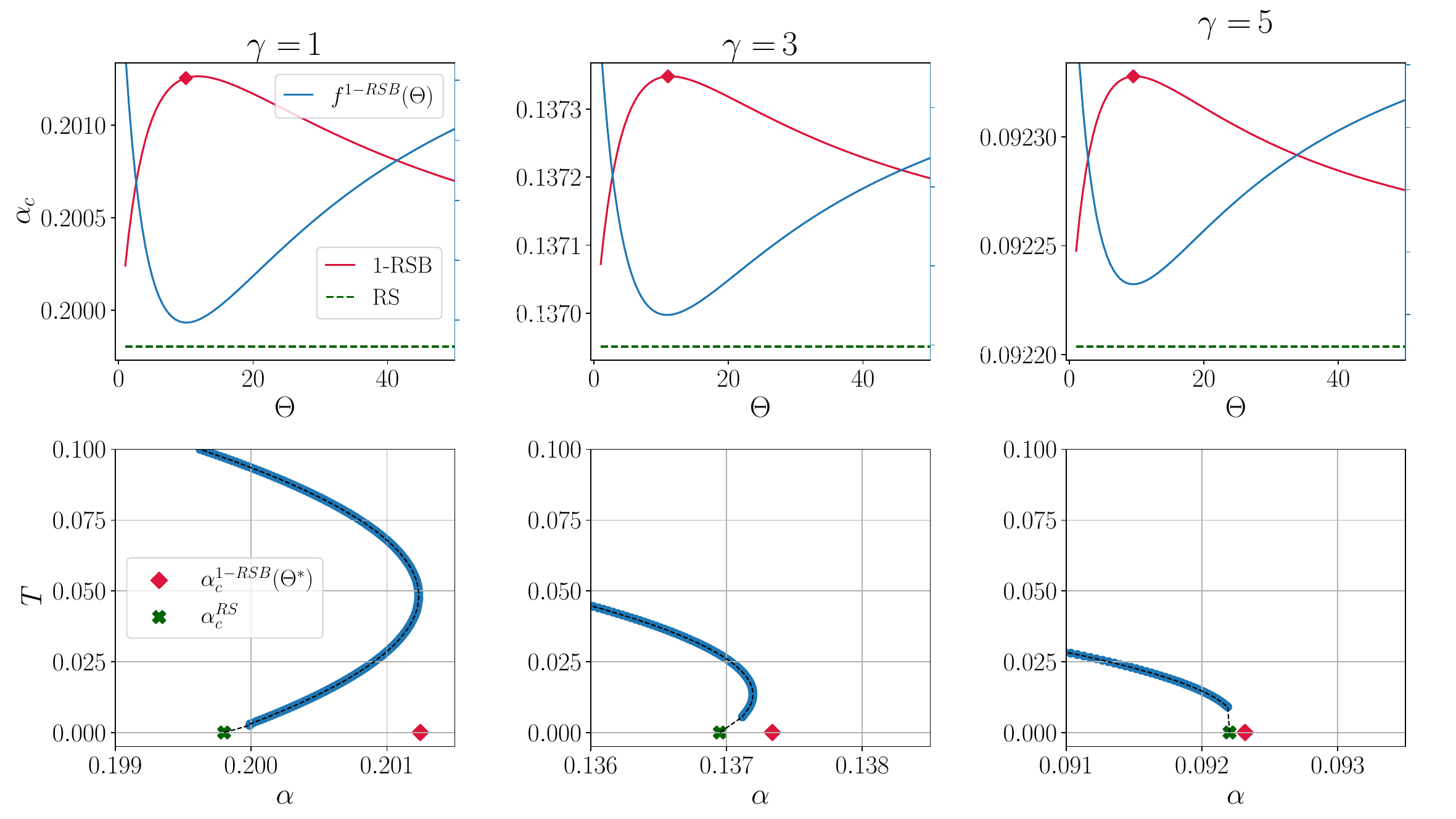}
\put(3,52.8){{\figpanel{a1}}}
\put(36,52.8){{\figpanel{a2}}}
\put(69,52.8){{\figpanel{a3}}}
\put(3,27.5){{\figpanel{b1}}}
\put(36,27.5){{\figpanel{b2}}}
\put(69,27.5){{\figpanel{b3}}}
\end{overpic}

\caption{The upper panels show the critical load $\alpha_c$ estimated through the 1-RSB self-consistent equations \eqref{:selfconsistent1RSBT0} at fixed $\Theta$ and plotted as a function of the latter; we superimpose the free energy of the retrieval state at the critical point (blue lines); for comparison, dashed green lines report the RS critical capacity computed in the previous section (see Fig. \ref{fig:phasediagram_T0}.). 
The bottom panels show a zoom of the BAM phase diagram at low temperature close to the MR-SG transition, where the re-entrance of the critical line is manifest. Scatter points identify the RS critical capacity (green) obtained as a continuation of the RS spinodal line (blue points); red dots identify the critical capacity obtained at the optimal value $\Theta^*$ minimizing the free energy of the corresponding upper panel.
Each column corresponds to one value of $\gamma$: \figpanel{a1}-\figpanel{b1}: $\gamma=1$; \figpanel{a2}-\figpanel{b2}: $\gamma=3$; \figpanel{a3}-\figpanel{b3}: $\gamma=5$. \label{fig:RSB}}
\end{figure}
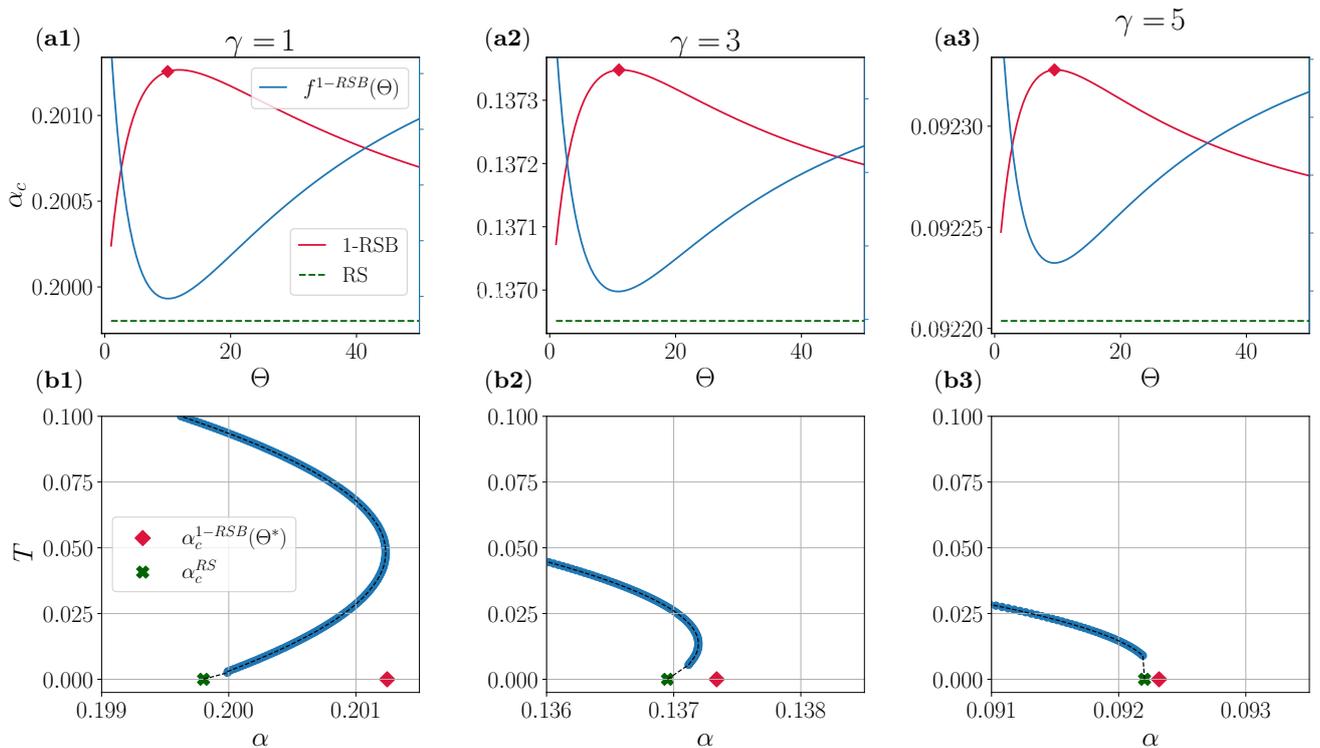

Instead of solving numerically Eqs. \eqref{eq:selfconsistent1RSB} at finite temperature we just focus on the $T\to 0$ limit, with the main goal of determining if the critical capacity increases with respect to the RS estimation, thus reducing the re-entrant behaviour of the phase diagram at low temperatures. The analytic limit of the free energy can be easily computed by following the same kind of calculations as in \citep{crisanti1986saturation,AlbaneseDAM}. As discussed in the previous section, the two largest spin overlaps $Q_2, \bar{Q}_2$ tend to $1$ in such a way that the quantities $\chi = \beta \left(1-Q_2 \right)$, $\bar{\chi} = \beta \left(1-\bar{Q}_2 \right)$ have a finite limit when $\beta \to \infty$. In addition, the parameter $\theta $ is rescaled by $\beta$ so that $\Theta=\beta \theta$, as pointed out in \citep{Steffan1994rsb, crisanti1986saturation}.  With these rescalings, and defining for convenience the overlap differences $\delta Q= 1- Q_1$ and $\delta \bar{Q} = 1- \bar{Q}_1$, the inner expectations in Eqs. \eqref{eq:1RSBfreeen}-\eqref{eq:selfconsistent1RSB} (denoted with $\mathbb{E}_z$), can be performed analytically; further simplifying by considering just one pattern pair to be retrieved (i.e. $l=1$ as in Section \ref{subsec:RSphasediagramT0}), the $1-$RSB free energy in the $T\to 0$ limit reads 
\begin{align}
f\left(\alpha,\gamma\right)  = \; &  M\bar{M}+\frac{\alpha}{2}\left(P_{2}\chi+\bar{P}_{2}\bar{\chi}\right)+\frac{\alpha}{2}\Theta\left(P_{1} \delta  Q +\bar{P}_{1} \delta\bar{Q}\right) +\frac{\alpha}{2\Theta}\log\frac{\Delta_{\Theta}}{\Delta}-\frac{\alpha}{2\Delta_{\Theta}}\left[\left(1-\delta\bar{Q}\right)\kappa+\left(1-\delta Q\right)\bar{\kappa}\right] + \nonumber \\
 & -\frac{\gamma}{\Theta}\mathbb{E}_{\tau}\log \mathcal{G}_{+}\left(\tau\right)-\frac{\bar{\gamma}}{\Theta}\mathbb{E}_{\tau}\log \bar{\mathcal{G}}_{+}\left(\tau\right) + \frac{\gamma + \bar{\gamma}}{\Theta}\log 2\label{eq:1rsbFreeenT0}
 \end{align} where $\Delta=1-\chi \bar{\chi}$, $\Delta_{\Theta}$ is the analogous limit at $T=0$ of Eq. \eqref{eq:deltatheta1RSB} and we defined the following quantities:
\begin{subequations}
\begin{alignat}{2}
\mathcal{G}_{\pm}\left(\tau \right) &= e^{b\Theta}\left(1+\text{erf }\omega_{+}\right)\pm e^{-b\Theta}\left(1+\text{erf }\omega_{-}\right)\, , \qquad \qquad \qquad &\bar{\mathcal{G}}_{\pm}\left(\tau \right) & = e^{\bar{b}\Theta}\left(1+\text{erf }\bar{\omega}_{+}\right)\pm e^{-\bar{b}\Theta}\left(1+\text{erf }\bar{\omega}_{-}\right)\, , \\ 
a & =\sqrt{\bar{\gamma}\alpha\left(P_{2}-P_{1}\right)}\, , & \bar{a} &=\sqrt{\gamma\alpha\left(\bar{P}_{2}-\bar{P}_{1}\right)}\, , \\
b\left(\tau \right) & =\bar{\gamma}\bar{M}+\sqrt{\bar{\gamma}\alpha P_{1}}\tau, &\bar{b} \left(\tau \right)&= \gamma M +\sqrt{\gamma\alpha\bar{P}_{1}}\tau\, , \label{eq:bss}\\
\omega_{\pm} \left(\tau \right)& =\frac{\Theta a}{\sqrt{2}}\pm\frac{b}{\sqrt{2}a}, & \bar{\omega}_{\pm}\left(\tau \right)&=\frac{\Theta\bar{a}}{\sqrt{2}}\pm\frac{\bar{b}}{\sqrt{2}\bar{a}}\, . \label{eq:omegas}
\end{alignat} \label{subeq:Gandallparams}
\end{subequations}
Note that the functions $\mathcal{G}, \bar{\mathcal{G}}$ explicitly depend on $\tau$ through $\omega_{\pm},b$ and $\bar{\omega}_{\pm},\bar{b}$ respectively (Eqs. \eqref{eq:bss}-\eqref{eq:omegas}), but for simplicity we dropped all the dependencies at the r.h.s. of Eqs \eqref{subeq:Gandallparams}.
The self-consistent equations can be either obtained by imposing stationarity of Eq. \eqref{eq:1rsbFreeenT0} w.r.t. the $M,\bar{M},\chi, \bar{\chi}, \delta Q, \delta \bar{Q}$ or by directly taking the $T\to 0$ limit of Eqs. \eqref{:selfconsistent1RSBT0}. In either case, the final result is given by
\begin{subequations}
\begin{alignat}{2}
M & =\mathbb{E}_{\tau}\frac{\mathcal{G}_{-}\left(\tau\right)}{\mathcal{G}_{+}\left(\tau\right)}\, ,\qquad\qquad\qquad\quad\quad\qquad \qquad \qquad\qquad\qquad & \bar{M} & =\mathbb{E}_{\tau}\frac{\bar{\mathcal{G}}_{-}\left(\tau\right)}{\mathcal{\bar{G}}_{+}\left(\tau\right)}\, ,\\
\delta Q & =4\,\mathbb{E}_{\tau}\frac{\left(1+\text{erf}\,\omega_{+}\right)\left(1+\text{erf}\,\omega_{-}\right)}{\mathcal{G}_{+}^{2}\left(\tau\right)}\, ,& \delta\bar{Q} & =4\,\mathbb{E}_{\tau}\frac{\left(1+\text{erf}\,\bar{\omega}_{+}\right)\left(1+\text{erf}\,\bar{\omega}_{-}\right)}{\bar{\mathcal{G}}_{+}^{2}\left(\tau\right)}\, ,\\
\chi & =\sqrt{\frac{8}{\pi}}\frac{1}{a}e^{-\frac{1}{2}a^{2}\Theta^{2}}\mathbb{E}_{\tau}\frac{e^{-\frac{b^{2}}{2a^{2}}}}{\mathcal{G}_{+}\left(\tau\right)}\, ,& \bar{\chi}& =\sqrt{\frac{8}{\pi}}\frac{1}{\bar{a}}e^{-\frac{1}{2}\bar{a}^{2}\Theta^{2}}\mathbb{E}_{\tau}\frac{e^{-\frac{\bar{b}^{2}}{2\bar{a}^{2}}}}{\mathcal{\bar{G}}_{+}\left(\tau\right)}\, ,\\
P_{1} & =\Delta_{\Theta}^{-2}\left[1-\delta\bar{Q}+\bar{\kappa}^{2}\left(1-\delta Q\right)\right]\, ,& \bar{P}_{1} & =\Delta_{\Theta}^{-2}\left[1-\delta Q+\kappa^{2}\left(1-\delta\bar{Q}\right)\right]\, ,\\
P_{2} & =P_{1}+\left(\Delta\Delta_{\Theta}\right)^{-1}\left(\delta\bar{Q}+\bar{\kappa}\bar{\chi} \delta Q \right)\, ,& \bar{P}_{2} &=\bar{P}_{1}+\left(\Delta\Delta_{\Theta}\right)^{-1}\left(\delta Q +\kappa\chi \delta\bar{Q} \right)\, ,
\end{alignat}\label{:selfconsistent1RSBT0}
\end{subequations}
The numerical solution of the above set of equations allows us to determine the location of the critical load for a specific value of $\gamma$. In principle, an additional equation must be solved together with Eqs. \eqref{:selfconsistent1RSBT0} regarding the stationarity of $f$ w.r.t. the rescaled Parisi parameter $\Theta$. However, numerical solutions involving also the latter are typically more involved from the point of view of numerical stability. We instead proceed by fixing a certain value $\Theta$ and solving Eqs. \eqref{:selfconsistent1RSBT0}; then, the optimal value of $\Theta$ is determined \textit{a-posteriori} by looking at the minimum of the free energy \eqref{eq:1rsbFreeenT0} w.r.t. $\Theta$. A summary set of results is shown in Figure \ref{fig:RSB}: the upper panels display the 1-RSB critical load computed as a function of $\Theta$ (one column correspond to a fixed value of $\gamma$), and the corresponding free energy of the Retrieval state at the critical point; the lower panels show the BAM's phase diagram (as in Figure \ref{fig:Phasediagrams_all}) in the region close to the MR-SG spinodal, where we highlight the new position of the critical capacity as estimated through the 1RSB equations. We first notice that 1RSB critical capacity is always higher than the corresponding RS estimation (green dashed lines in the upper panels of Fig. \ref{fig:RSB}), and the latter is recovered in the two limits $\Theta\to \{0,\infty \}$. However, at a difference w.r.t. what discussed in \citep{Steffan1994rsb}, the optimal (w.r.t. $\Theta$) critical load \footnote{Note that the optimal value of the critical load w.r.t. $\Theta$ does not necessarily coincide with its maximum value, i.e. $ \alpha_c \left(\text{arg} \min_{\Theta} f(\Theta) \right) \neq \max_{\Theta} \alpha_c\left(\Theta\right)$} slight exceeds the maximum value attained at finite $T$. In any case, the difference between the 1-RSB and the RS critical loads are of order $O\left(10^{-3}-10^{-4}\right)$ so the difference between the two estimations are in any case in agreement with the simulation results discussed in the next section. As a final remark, note that the re-entrant behavior of the phase diagram reduces as the asymmetry between the two layers increases: this is also confirmed by the optimal value $\Theta^*$ at the critical point, which numerically turns out to be a decreasing function of $\gamma \lessgtr 1$. In other words, the temperature at which the RS instability occurs for the retrieval solution shifts to lower temperatures.

\section{Numerical simulations \label{sec:numericalsim}}
In this section we provide a numerical characterization of the BAM's dynamics, with the aim of numerically determining the retrieval-spin glass spinodal point. At $T=0$, the dynamic rules for updating a single neuron belonging to one of the two layers are given by:
\begin{equation}
\sigma_{i}^{t+1} =\text{sign}\left(\sum_{j}w_{ij}\bar{\sigma}_{j}^{t} \right), \qquad \qquad \qquad
\bar{\sigma}_{j}^{t+1}  =\text{sign}\left(\sum_{i}w_{ij}\sigma_{i}^{t} \right),\label{eq:dynseq_sigmaboth}
\end{equation}
respectively if the chosen neuron belongs to layer $1$ or layer $2$. Eq. \eqref{eq:dynseq_sigmaboth} represents a noiseless version of the Glauber's sequential dynamics for the BAM. At finite $\beta$, the state of each neuron is sampled from a single-variable distribution, uniquely defined by a local field. In particular, for spins belonging to either layer, these probabilities are given by:
\begin{align}
    \sigma_{i}^{t+1} & \sim \frac{e^{\beta h_i \left(\bar{\boldsymbol{\sigma}}^{t} \right) \sigma_i^{t+1}}}{2 \cosh \beta h_i \left(\bar{\boldsymbol{\sigma}}^t \right)},\qquad \qquad \qquad  h_i \left(\bar{\boldsymbol{\sigma}} \right) = \sum_{j}w_{ij}\bar{\sigma}_{j}, \label{eq:dyn_beta_sigma_i_tp1}\\
    \bar{\sigma}_{j}^{t+1} & \sim \frac{e^{\beta h_j \left(\boldsymbol{\sigma}^{t} \right) \bar{\sigma}_j^{t+1}}}{2 \cosh \beta h_j \left(\boldsymbol{\sigma}^t \right)},\qquad \qquad  \qquad h_j \left(\boldsymbol{\sigma} \right) = \sum_{i}w_{ij}\sigma_{i}.
    \label{eq:dyn_beta_sigmabar_j_tp1}
\end{align}
It is easy to show how the above update rules satisfy detailed balance at finite $\beta$ and the corresponding equilibrium distribution is given by the Boltzmann probability with the BAM's Hamiltonian \eqref{eq:hamiltonian_starting}:
indeed, it is sufficient to rewrite the original model as a generalized
Hopfield with $N+{\N}$ neurons, having the following synaptic matrix
\begin{equation}
\boldsymbol{J}=\begin{pmatrix}\boldsymbol{0} & \boldsymbol{W}\\
\boldsymbol{W}^{T} & \boldsymbol{0}
\end{pmatrix},\label{eq:synaptic_matrix_BAM_asHopfield}
\end{equation}
where $\boldsymbol{W}$ is the BAM's
weight matrix defined in Eq. \eqref{eq:coupling_matrix}. 
In the following, we will focus on the retrieval of pure states (i.e. of a single pattern pair) by storing an increasing number of patterns in the network to characterize the first-order spinodal between the retrieval and the spin-glass phases.
To study the retrieval properties of the model, the initial condition on at least one of the two layers must have a high overlap with the pattern to be retrieved: otherwise, the dynamics gets stuck in spurious attractors reminiscent of the spin-glass fixed point (which, as mentioned earlier, also exists in the retrieval phase) having zero overlap with any pattern. Therefore, the dynamics starts by choosing an initial configuration that is close enough to a pattern pair $\mu$ (in what follows we assume without loss of generality $\mu=1$), by random flipping each neuron with probability $\varepsilon\in\left(0,1\slash2\right)$ w.r.t the pattern component. As a consequence, the initial Mattis magnetization over a layer is equal (on average) to  $M^{\left(t=0\right)}=1-2\varepsilon$. In principle, since there are two layers in the BAM, we use $\varepsilon_{1}$ (or $\varepsilon_{2}$) to indicate the initial noise on the first (or second) layer: as will become clear below, choosing different values for the initial noise on the largest/smallest layer have strongly different outcomes if the BAM is asymmetric.

Rather than using a random sequential update scheme at each iteration of the dynamics, the bipartite structure of the BAM naturally suggests to simultaneously update the spins in one layer given the entire state of the other layer: such a ``parallel'' update exploits the reverberation mechanism in which information on one layer is transferred to the other at once, as originally described in Ref. \citep{kosko_bidirectional_1988}: for example, note how the local fields in Eqs. \eqref{eq:dyn_beta_sigma_i_tp1}-\eqref{eq:dyn_beta_sigmabar_j_tp1} acting on each neuron in one layer depend exclusively on the neuron states of the other layer, again by virtue the bipartite structure of the model. This choice also allows for a significant speedup of simulations (especially on GPU architectures), as shown by the typical sizes that can be simulated in a reasonable amount of time.
Therefore, all the results presented in the current section are obtained using a simultaneous update of Eq. \eqref{eq:dyn_beta_sigma_i_tp1} for all the neurons in layer $1$ and Eq. \eqref{eq:dyn_beta_sigmabar_j_tp1} for all the neurons in layer $2$, in alternate sequence, for a total of $N_{\text{steps}}$. Without loss of generality, we always start the dynamics by updating layer $2$, so that the only relevant initial condition is the starting configuration on layer $1$ (eventually, with an initial noise $\varepsilon_1$). For completeness, in Appendix \ref{sec:appendix_seqdyn} we also report an equivalent set of results using a random sequential update of Eqs. \eqref{eq:dyn_beta_sigma_i_tp1}-\eqref{eq:dyn_beta_sigmabar_j_tp1} (using smaller system sizes), which display an identical behavior.

\begin{figure}
\begin{overpic}[width=\textwidth]{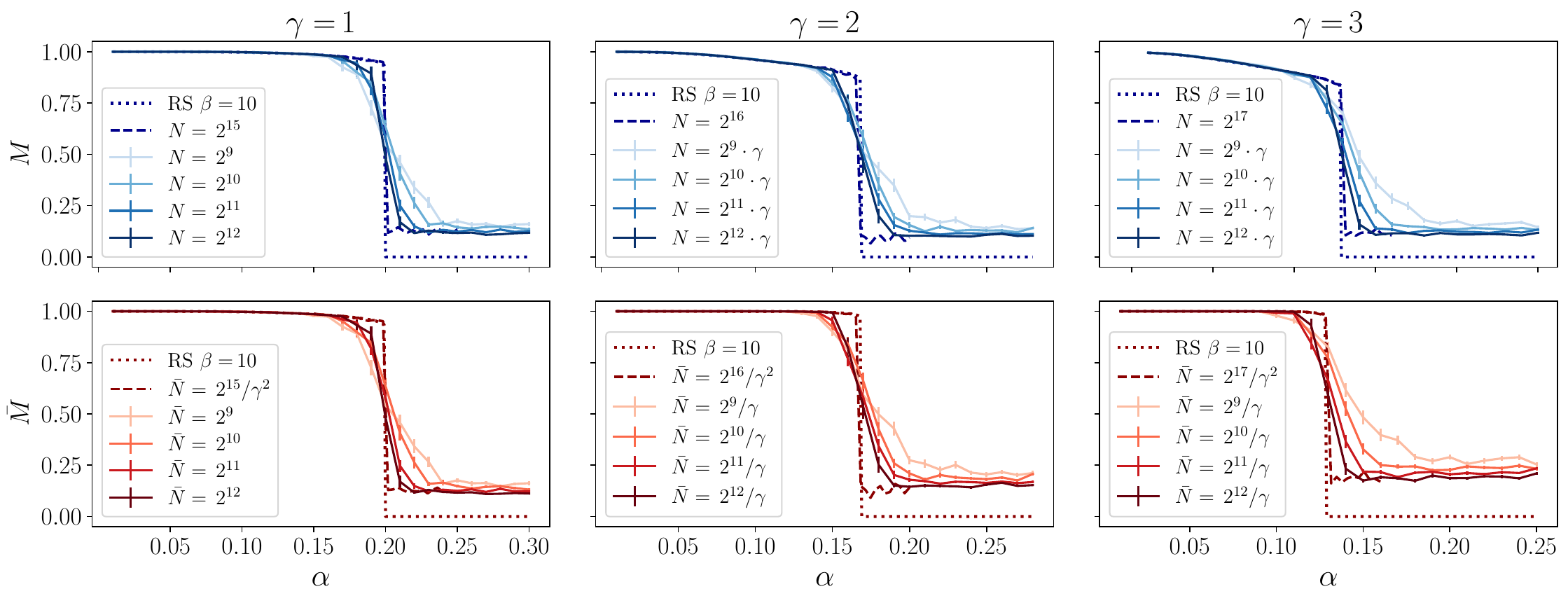}
\put(31,33.8){{\figpanel{a1}}}
\put(63,33.8){{\figpanel{a2}}}
\put(95,33.8){{\figpanel{a3}}}
\put(31,17){{\figpanel{b1}}}
\put(63,17){{\figpanel{b2}}}
\put(95,17.){{\figpanel{b3}}}
\end{overpic}
\caption{Numerical estimation of the retrieval/spin-glass phase transition. Simulations are performed using a parallel dynamics at $\beta=10$ with $N_{\text{steps}} = 5 \times 10^4$ for $3$ values of $\gamma$. \figpanel{a1}-\figpanel{b1}: $\gamma=1$; \figpanel{a2}-\figpanel{b2}: $\gamma=2$;
\figpanel{a3}-\figpanel{b3}: $\gamma=3$;
Each panel shows the final Mattis magnetization over each layer w.r.t. to the pattern to be retrieved. In all cases, the dynamics starts from the pattern to be retrieved with a small noise $\varepsilon_{1}=0.1$ on layer $1$. Results are plotted as a function of $\alpha$ for four different system sizes (shown in the caption) and averaged over $100$ different realizations of the disorder. Dashed lines in each panel show the results for a single instance with a very large system size. Dotted lines in each panel denote the RS saddle point solution at $\beta=10$ (i.e. a horizontal cut in the phase diagrams of Fig. \ref{fig:Phasediagrams_all}).  \label{fig:storage_pardynamics}} 
\end{figure}
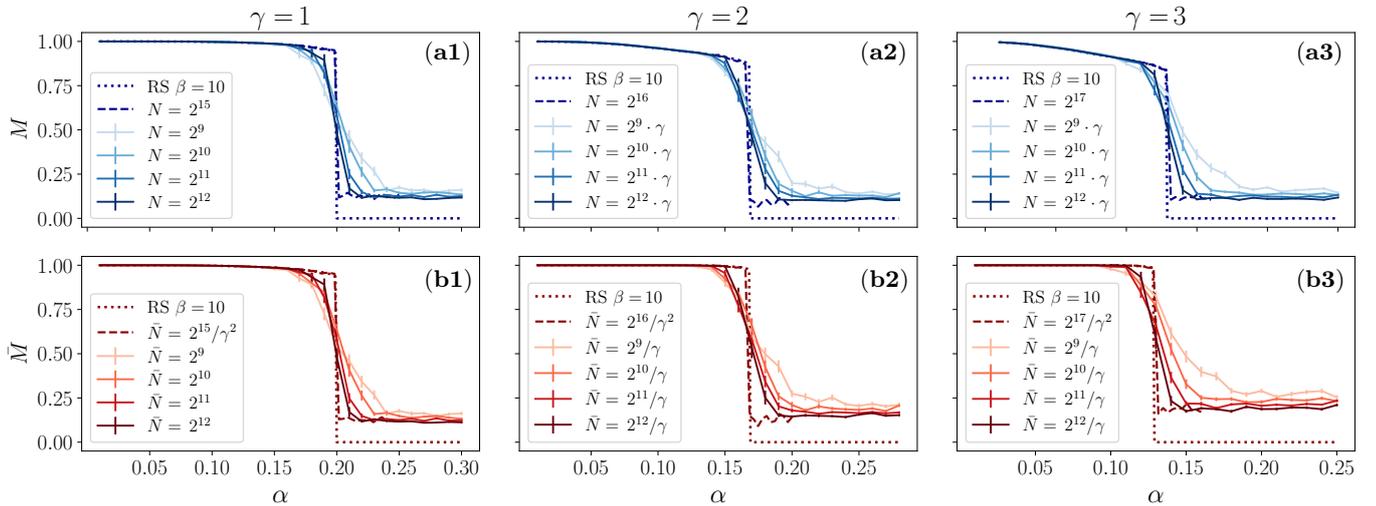

A summary of numerical results is shown in Figure \ref{fig:storage_pardynamics}, considering three values of the asymmetry, i.e. $\gamma=1,2,3$ (each value corresponds to one column). Each panel shows the final overlap of spin configurations calculated at the end of the simulation with respect to the original pattern (i.e., the Mattis magnetizations \eqref{eq:mattis_overlap_starting}), on layer 1 (top panels) and 2 (bottom panels), plotted as a function of $\alpha$.
We find an overall good agreement between the numerical simulations in predicting the location of the transition between retrieval/spin glass, compared to the RS theory (plotted in dotted lines). Moreover, the empirical transition curves become sharper as the total system size $L$ is increased, as expected for finite sizes. Figure \ref{fig:storage_pardynamics} also reports the results for a single large system size, confirming the self-averaging property of the free energy.

In Figure \ref{fig:storage_pardynamics} we further notice how, even beyond the predicted critical capacity, the magnetization does not fall to $0$: a first explanation is that finite-size effect can result into the dynamics being stuck in spurious attractors characterized by a partial overlap with respect to more than one pattern (mixed states), that are not present in the MF theory in this regime, a phenomenon already observed in \citep{leuzzi_quantitative_2022} for the Hopfield model. Secondly, this phenomenon can be understood also in terms of the lack of patterns' orthogonality at finite size, and it would would explain why in both Figure \ref{fig:storage_pardynamics} \figpanel{a2}-\figpanel{b2}-\figpanel{a3}-\figpanel{b3} (i.e. in the asymmetric scenarios with $\gamma=2,3$) the final magnetizations on the smallest layer are slightly higher than the largest layer's ones even at $\alpha> \alpha_c$.

\begin{figure}
\begin{overpic}[width=\textwidth]{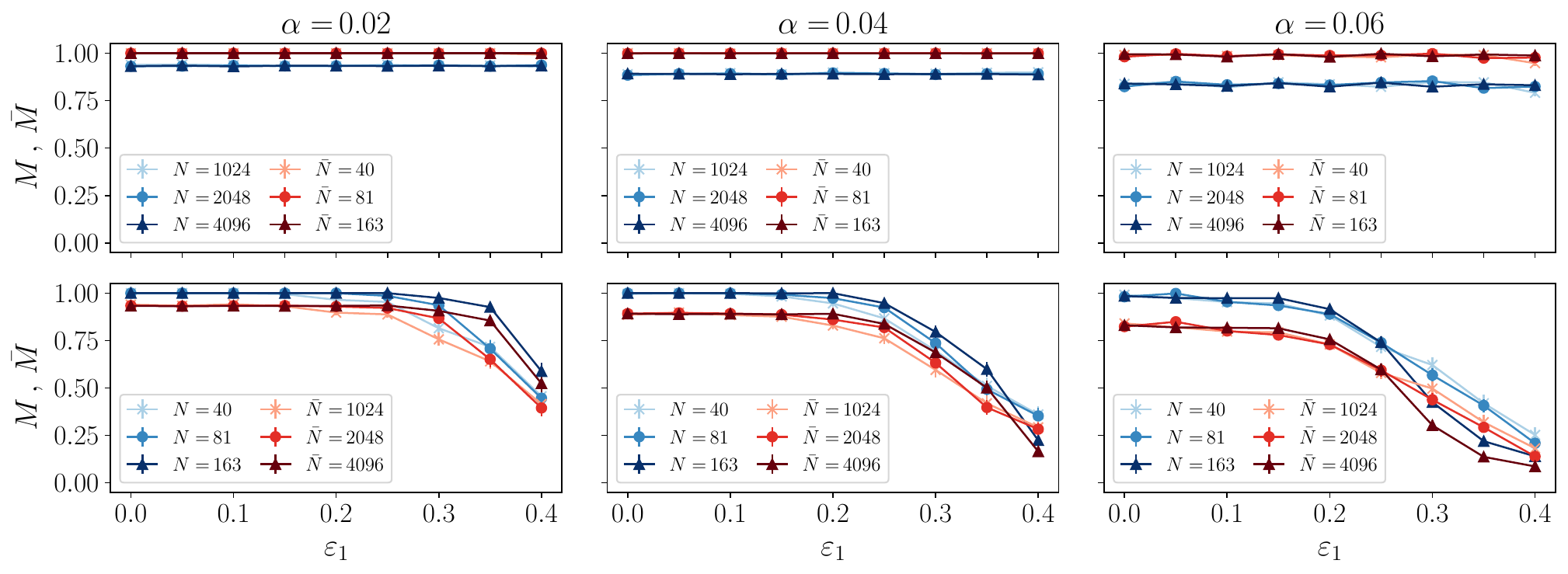}
\put(8,28){{\figpanel{a1}}}
\put(39.5,28){{\figpanel{a2}}}
\put(71,28){{\figpanel{a3}}}
\put(8,12.7){{\figpanel{b1}}}
\put(39.5,12.7){{\figpanel{b2}}}
\put(71,12.7){{\figpanel{b3}}}
\end{overpic}
\caption{Basin of attraction in the asymmetric BAM. Simulations performed using a parallel dynamics at $\beta = 10$ for $N_{\text{steps}}=10^4$. The upper row (panels \figpanel{a1}$\to$\figpanel{a3}) shows results at $\gamma = 5$, so that the first layer is the largest one (i.e. $N> {\N} $). The bottom row (panels \figpanel{b1}$\to$\figpanel{b3}) refers to the opposite scenario where the two layers are swapped, namely $\gamma= 1 \slash 5$ and $N < {\N}$.  Each panel shows the final Mattis magnetization of the two layers $M,\bar{M}$ (in blue/red shades respectively), plotted as functions of the initial noise on layer $1$ (denoted with $\varepsilon_{1}$), for three different system sizes  $L\in \left\{ 2^{10}\slash \gamma, 2^{11}\slash \gamma,2^{12}\slash \gamma \right\}$. Each datapoint represents the average over $100$ realizations of the disorder. Each column refers instead to a different value of the network load $\alpha<\alpha_c\left(\gamma=5 \right)$. \label{fig:basin}}

\end{figure}
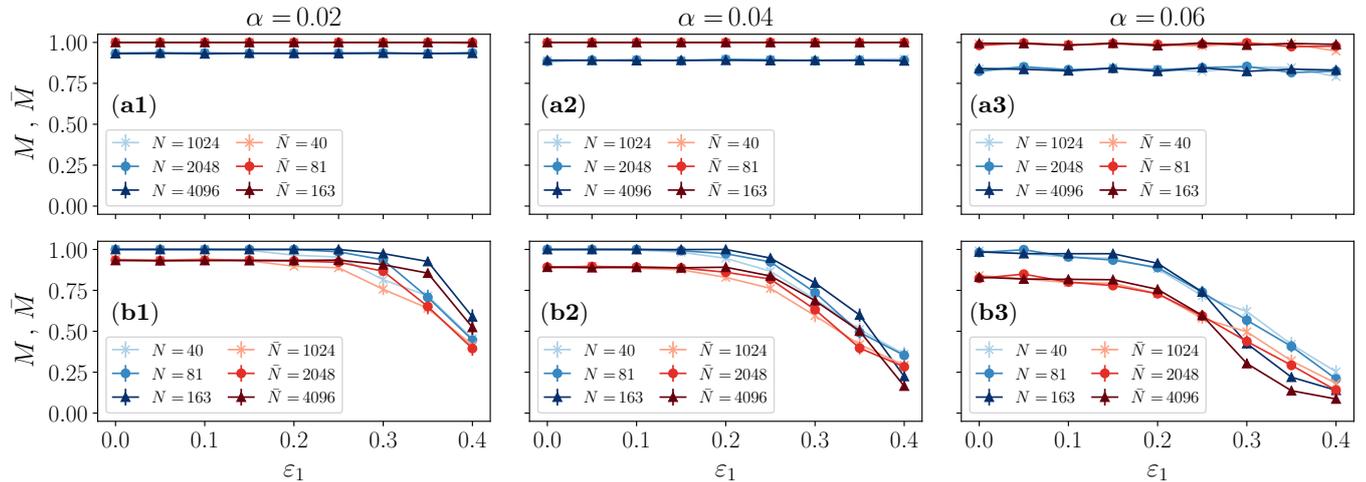

\paragraph{Basin of attraction}
Finally, we report a numerical experiment to quantify the different basins of attraction of the two layers in the asymmetric BAM (i.e $\gamma \neq 1$). A summary set of results is shown in Figure \ref{fig:basin}, where we performed simulations on a BAM with increasing size $L$ while keeping fixed the asymmetry between the two layers (in all the results of Fig. \ref{fig:basin}, $\gamma$ is equal either to $5$ or $1 \slash 5$). The network load $\alpha$ is chosen in such a way that the network is expected to be inside the retrieval phase within the MF theory, so that $\alpha < \alpha_c (\gamma=5) \approx 0.092$. 
The initial noise (varying on the horizontal scale in all the panels in Fig. \ref{fig:basin}) is always set on the first layer ($\varepsilon_1$). What changes between the upper and bottom panels is that, in the former, the first layer is the largest one ($\gamma=5$), while in the latter the situation is swapped ($\gamma=1\slash 5$).
Therefore, when the first layer is the largest one, even a strong initial noise does not affect the retrieval capabilities of the machine: in particular, the values plotted in all the upper panels of the final Mattis magnetizations (on both layers) are consistent with the MF prediction. On the other hand, in the bottom panels the two layers are inverted, so that the dynamics is initialized with an increasing noise on the smallest layer. As a consequence, by increasing such noise the BAM is not anymore able to retrieve the pattern pairs on the two layers, as evidenced by the decreasing Mattis magnetizations in Fig. \ref{fig:basin} \figpanel{b1}-\figpanel{b2}-\figpanel{b3}. This is consistent with the original claim in Ref. \citep{kurchan_statistical_1994} that the smallest layer has a smaller basin of attraction.

\section{Understanding BAM's retrieval mechanism: analogy with coupled Hopfield models in the low-load regime \label{sec:Pattern-retrieval-in}}

In this section, you will find a heuristic discussion to better understand how retrieval is attained in the BAM. 
The main differences between the Hopfield model and the BAM is that in the latter, the patterns retrieved in both layers correspond to two different sets of variables: $\bm{\xi}$ and $\bm{\bar{\xi}}$. This is clear from the analytical formulation of the two models: In the Hopfield model, the Hamiltonian can be expressed as the square of a Mattis magnetization, whereas in the BAM, it is the product over the Mattis magnetization over each layer. In fact, we can observe that the Hopfield potential is a quadratic function of the Mattis magnetization: $m^2(\boldsymbol{\sigma})$, hence being stable. While, the BAM Hamiltonian resembles a saddle given by the product of the Mattis magnetization between the two different layers: $m(\boldsymbol{\sigma})\bar{m}(\boldsymbol{\bar{\sigma}})$: the main difference is that, in the BAM, we cannot retrieve a certain pattern on both layers if the two magnetizations have opposite signs. This structure naturally suggests that we can interpolate between the two situations to construct some sort of generalization of the Hopfield model with tunable intra/inter synaptic connections, or to put it differently, to move from a quadratic potential to a one with unstable directions.

Consequently, the BAM can be viewed as a generalized Hopfield model over the union set of the neurons of the two layers $\boldsymbol{\sigma}\cup\bar{\boldsymbol{\sigma}}$, where only the intersynaptic connections between the two subsets are maintained, while all intrasynaptic connections are set to $0$ (see Eq. \eqref{eq:synaptic_matrix_BAM_asHopfield}).
This mechanism allows for the BAM to retrieve only associated information
on each layer: namely, if the pattern $\boldsymbol{\xi}^{\mu}$ is recalled
on layer $1$, layer $2$ will recall its corresponding "dual" pattern
$\bar{\boldsymbol{\xi}}^{\mu}$. 
In other words, the BAM is able to retrieve a pattern represented by $\boldsymbol{\xi}^{\mu} \cup \bar{\boldsymbol{\xi}}^{\mu}$ (for the set $\boldsymbol{\sigma}\cup\bar{\boldsymbol{\sigma}}$ of neurons), with a reduced number of interaction weights, $N \times {\N}$: a number to be compared with the $\!\sim\!\left(N \!+\! {\N}\right)^2$ weights that would be needed in the equivalent Hopfield case.

For the following analysis, we consider a generalization of the starting
Hamiltonian \eqref{eq:hamiltonian_starting}, introducing an additional
parameter $\tau$:
\begin{equation}
H_{\tau}\left(\boldsymbol{\sigma}, \bar{\boldsymbol{\sigma}}\right)\!=\!-\!\sum_{\mu=1}^{K}\left[\frac{\left(1\!-\!\tau\right)}{2}\left(\frac{1}{\sqrt{N}}\sum_{i}\xi_{i}^{\mu}\sigma_{i}\right)^{2}\!+\!\frac{\left(1\!-\!\tau\right)}{2}\left(\frac{1}{\sqrt{{\N}}}\sum_{j}\bar{\xi}_{j}^{\mu}\bar{\sigma}_{j}\right)^{2}\!+\!\tau\left(\frac{1}{\sqrt{N}}\sum_{i}\xi_{i}^{\mu}\sigma_{i}\right)\left(\frac{1}{\sqrt{{\N}}}\sum_{j}\bar{\xi}_{j}^{\mu}\bar{\sigma}_{j}\right)\right].\label{eq:hamiltonianinterp_tau}
\end{equation}
Note that a similar kind of interpolation has been employed to overcome the obstacle of the non convexity of the interaction for the free energy of layered spin-glass models \citep{Alberici2020,Alberici2021}.
Clearly, Eq. \eqref{eq:hamiltonianinterp_tau} recovers the original BAM Hamiltonian when $\tau=1$. On the other hand, when $\tau=0$, the generalized Hamiltonian describes two independent Hopfield models with $N$ and ${\N}$ units, respectively, each of which has its own set of stored patterns in its synaptic matrix (which is fully connected within each layer), i.e.
\begin{equation}
H_{\tau=0}\left(\boldsymbol{\sigma}, \bar{\boldsymbol{\sigma}}\right)=-\frac{1}{2}\sum_{\mu=1}^{K}\left(\frac{1}{\sqrt{N}}\sum_{i}\xi_{i}^{\mu}\sigma_{i}\right)^{2}-\frac{1}{2}\sum_{\mu=1}^{K}\left(\frac{1}{\sqrt{{\N}}}\sum_{j}\bar{\xi}_{j}^{\mu}\bar{\sigma}_{j}\right)^{2} = -\frac{N}{2} \sum_{\mu=1}^{K} \left(m^\mu(\boldsymbol{\sigma})\right)^2 - \frac{\bar{N}}{2} \sum_{\mu=1}^{K} \left(\bar{m}^\mu(\boldsymbol{\bar{\sigma}})\right)^2.\label{eq:limit_twoHop_tau0}
\end{equation}
Indeed, on each layer one can define a synaptic matrix constructed
via the usual Hebb's rule, namely,
\begin{equation}
J_{i_{1},i_{2}}=\frac{1}{N}\sum_{\mu=1}^{K}\xi_{i_{1}}^{\mu}\xi_{i_{2}}^{\mu},\qquad\bar{J}_{j_{1}j_{2}}=\frac{1}{{\N}}\sum_{\mu=1}^{K}\bar{\xi}_{j_{1}}^{\mu}\bar{\xi}_{j_{2}}^{\mu},\label{eq:jij_twohopfields}
\end{equation}
as it can be easily verified by expanding the squares in Eq. \eqref{eq:limit_twoHop_tau0}.
For a generic value of $\tau\in\left(0,1\right)$, Eq. \eqref{eq:hamiltonianinterp_tau}
describes a mixture of two Hopfield models with both intra-synaptic and inter-synaptic connections whose relative strength is tuned by the value of $\tau$. The main difference between the two limiting cases is that when $\tau=0$, the partition function of the model defined in Eq. \eqref{eq:hamiltonianinterp_tau},
factorizes completely over the two spin sets, which means that each Hopfield model by itself can retrieve any pattern, regardless on the other.
On the other hand, when $\tau=1$ only retrieval of corresponding couples of
patterns with the same index $\mu$ is possible. For this reason, the question naturally arises under what conditions the retrieval of individual information is possible as a function of $\tau$. The following analysis is performed in the regime of low load, i.e., when the number of patterns is
$K$ is finite relative to the number of spins on each layer, using the standard procedure used for the Hopfield model (e.g., as discussed in detail in \citep{coolen_statistical_2000}). For simplicity, we consider all patterns to be binary vectors with i.i.d. components drawn with equal probability between $\{-1,1\}$. Note, however, that retrieval in the low-load scenario should be robust w.r.t. the distribution chosen for the pattern components by generalizing recent arguments about the Hopfield model (see, e.g., Ref. \citep{barra_phase_2018}) 
In the low-load regime, the noise induced by the patterns is subextensive relative to the system size, so it is not necessary to introduce replicas or implement Guerra's interpolation technique to address this. The average quenched free energy can be easily calculated by a saddle point method after introducing the usual set of Mattis magnetizations on each layer. After some calculations, the density of the free energy corresponding to the Eq. \eqref{eq:hamiltonianinterp_tau} model can be written as
\begin{align}
f\left(\beta, \gamma, \tau \right) =& + \frac{1}{2}\gamma\left(1-\tau\right)\sum_{\mu}m_{\mu}^{2}-\frac{\gamma}{\beta}\mathbb{E}_{\bm{\xi}}\log\text{cosh}\left\{ \beta\sum_{\mu}\left[\left(1-\tau\right)m_{\mu}+\tau\bar{\gamma}\bar{m}_{\mu}\right]\xi^{\mu}\right\}\nonumber \\
 & +\frac{1}{2}\bar{\gamma}\left(1-\tau\right)\sum_{\mu}\bar{m}_{\mu}^{2}-\frac{\bar{\gamma}}{\beta}\mathbb{E}_{\bar{\bm{\xi}}}\log\text{cosh}\left\{\beta\sum_{\mu}\left[\left(1-\tau\right)\bar{m}_{\mu}+\tau\gamma m_{\mu}\right]\bar{\xi}^{\mu}\right\}\nonumber \\
 & +\tau\sum_{\mu}m_{\mu}\bar{m}_{\mu},\label{eq:free_energy_interp_tau}
\end{align}
where $m_{\mu},\bar{m}_{\mu}$ are defined as in Eqs. \eqref{eq:mattis_both}.
It is easy to check how the above expression recovers the BAM's free
energy in the low load limit -- i.e. \eqref{eq:RS_freeenergy} at $\alpha=0$
-- when $\tau=1$, and the sum of two Hopfield free energies (again,
in the low-load limit) - each rescaled by the shape parameter (resp.
$\gamma$ for layer $1$ and $\bar{\gamma}$ for layer $2$) - when
$\tau=0$.
The values of $\left\{ m_{\mu},\bar{m}_{\mu}\right\} _{\mu=1}^{K}$
can be found by imposing stationarity of \eqref{eq:free_energy_interp_tau},
which determines the following self-consistent equations:
\begin{align}
m_{\mu} & =\mathbb{E}_{\boldsymbol{\xi}}\xi^{\mu}\text{tanh}\left\{ \beta\sum_{\nu}\left[\left(1-\tau\right)m_{\nu}+\tau\bar{\gamma}\bar{m}_{\nu}\right]\xi^{\nu}\right\},\label{eq:m_saddlepoint_mu}\\
\bar{m}_{\mu} & =\mathbb{E}_{\bar{\boldsymbol{\xi}}}\bar{\xi}^{\mu}\text{tanh} \left\{ \beta\sum_{\nu}\left[\left(1-\tau\right)\bar{m}_{\nu}+\tau\gamma m_{\nu}\right]\bar{\xi}_{j}^{\nu} \right\}.\label{eq:mb_saddle_point_mu}
\end{align}
The equilibrium behavior of the model can again be studied by numerically solving the above equations. The simplest nontrivial scenario arises with $K=2$, so that the magnetization vectors have at most two nonzero components for each layer, parameterized as follows:
\begin{equation}
\boldsymbol{m} = \left(m_{1},m_2\right)^{T} \qquad \qquad \text{and} \qquad \qquad \bar{\boldsymbol{m}}  =\left(\bar{m}_{1},\bar{m}_{2}\right)^{T}.
\end{equation}

\begin{figure}
\begin{overpic}[width=\textwidth]{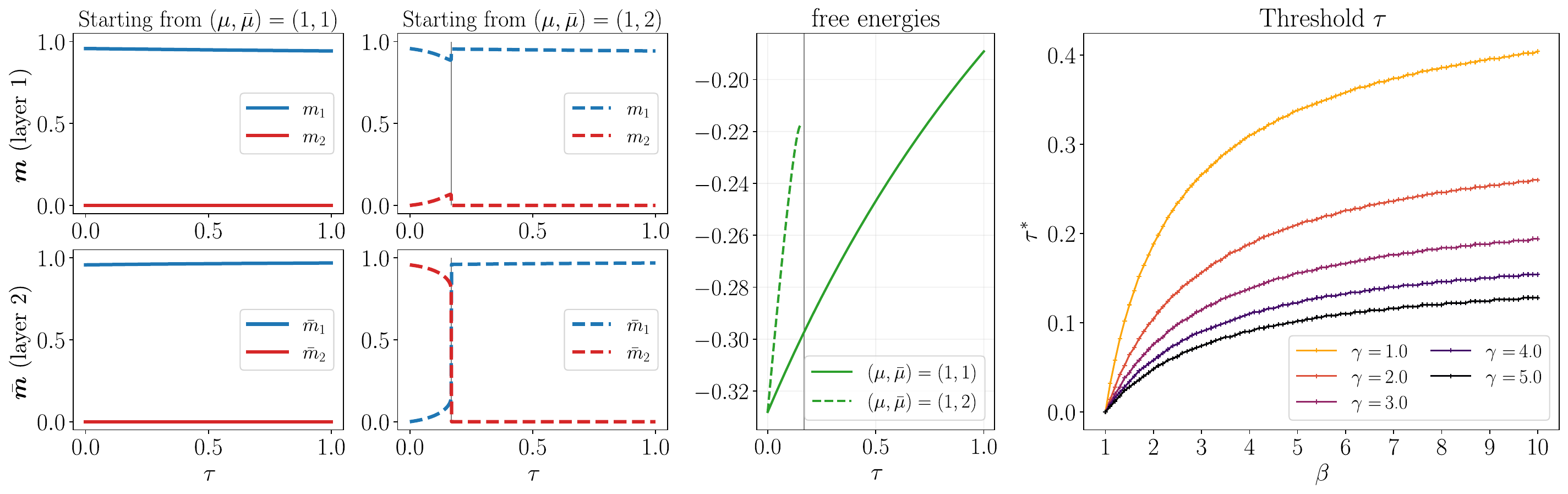}
\put(6,23){{\figpanel{a1}}}
\put(6,9){{\figpanel{a2}}}
\put(30,23){{\figpanel{a3}}}
\put(30,9){{\figpanel{a4}}}
\put(52,27){{\figpanel{a5}}}
\put(70,27){{\figpanel{b}}}
\end{overpic}

\caption{Panels labelled with letter \figpanel{a} show results on the saddle point equations \eqref{eq:m_saddlepoint_mu}-\eqref{eq:mb_saddle_point_mu} obtained varying $\tau$ for the retrieval of
the same/different pattern index on the two layers. \figpanel{a1}-\figpanel{a2}: Mattis magnetizations on 2 pattern components (shown resp. in blue and red solid lines and labelled with subscripts $1,2$) over layer 1 \figpanel{a1} and layer 2 \figpanel{a2}, obtained by initializing the saddle point equations with an initial high magnetization over pattern $1$ on both layers, i.e. $\mu,\bar{\mu} = 1$. \figpanel{a3}-\figpanel{a4} (again showing the Mattis magnetizations on layer 1 and 2 respectively) represent the same results but starting from a high magnetization on different patterns for the two layers: in particular, starting from an initial high overlap with pattern $1$ on the first layer and pattern $2$ on the second. Panel \figpanel{a5} shows the free energies of both solutions: the solid green line refers to the saddle points obtained in \figpanel{a1}-\figpanel{a2}, while the dashed line refers to \figpanel{a3}-\figpanel{a4}. All the results of \figpanel{a}-panels  are obtained at $\beta = 2$ and $\gamma = 1.1$, and plotted as a function of $\tau$. \figpanel{b}: Threshold value of $\tau$ after which retrieval of different pattern indexes in each of the layers is not possible, plotted as a function of $\beta>1$ for
different values of $\gamma$ (shown in the legend). \label{fig:sectionV}}
\end{figure}
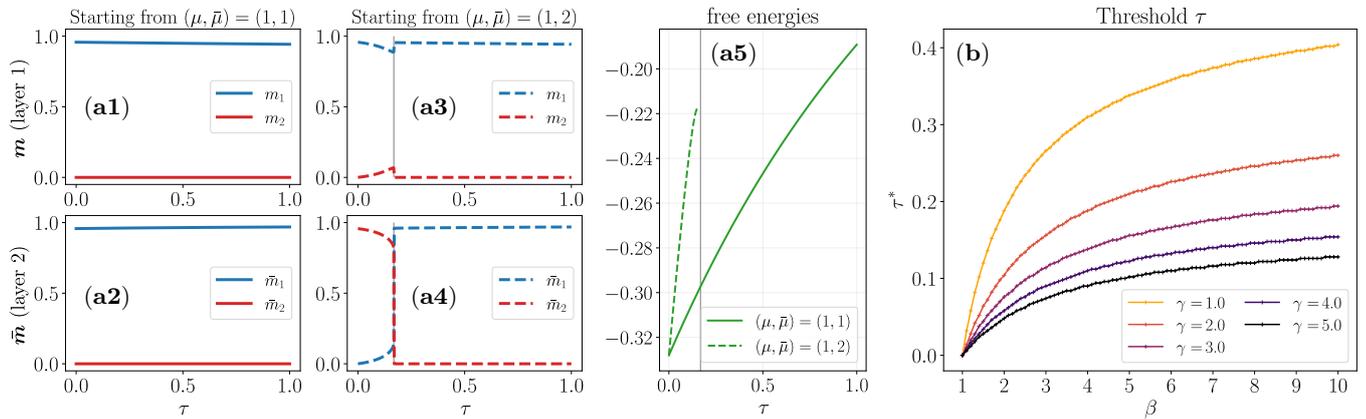

It is easy to check, both numerically and analytically, that independently
on the value of $\tau$ the only solution at $\beta<1$ is the paramagnetic
one, where all the magnetizations are null. At $\beta=1$. a ferromagnetic
solution appears, independently on the value of $\tau$.
It is perhaps more interesting to look at the numerical solutions by focusing on a certain value of $\beta>1$ (and $\gamma$) and analyze its behaviour as a function of $\tau$: an example set of results is
shown in Figure \ref{fig:sectionV}, panels \figpanel{a1}$\to$\figpanel{a5}, obtained at $\beta=2$ and $\gamma=1.1$ (although the qualitative
behaviour remains unchanged by lowering $\beta$ or changing $\gamma$). The four left panels \figpanel{a1}$\to$\figpanel{a4} show the two components of the Mattis magnetization for the two layers (respectively in
the upper and lower panel), obtained by solving Eqs. \eqref{eq:m_saddlepoint_mu}-\eqref{eq:mb_saddle_point_mu} at different values of $\tau$ (varying on the horizontal axis), and with respect to the two different patterns (with indexes $\mu,\bar{\mu}=1,2$). The left panels \figpanel{a1}-\figpanel{a2} refer to fixed points obtained by starting from a high magnetizations
on the \textit{same} pattern index on both layers, i.e. $\boldsymbol{m}^{0}=\left(1-\varepsilon,\varepsilon\right)^{T}$
and $\bar{\boldsymbol{m}}^{0}=\left(1-\bar{\varepsilon},\bar{\varepsilon}\right)^{T}$,
with $\varepsilon,\bar{\varepsilon}$ being arbitrary small. Such a ferromagnetic solution is always found at any value of $\tau$, and the corresponding
free energy is plotted in the panel \figpanel{a5} (full green line). We conclude that the retrieval of the same pattern is a stable fixed point in the full range $\tau \in [0,1]$.
Panels \figpanel{a3}-\figpanel{a4} show the same quantities, but this time the initial magnetization vector has a high value
on a \textit{different} pattern over each layer, e.g. $\boldsymbol{m}^{0}=\left(1-\varepsilon,\varepsilon \right)^{T}$
and $\bar{\boldsymbol{m}}^{0}=\left(\bar{\varepsilon},1-\bar{\varepsilon}\right)^{T}$. 
As clear from the figure, the retrieval of different patterns (in each of the layers) can be attained only at small values of $\tau$, i.e. when the two Hopfield models are almost independent and the strength of their inter-synaptic connections falls below a certain (temperature-dependent) threshold. Above such threshold - called $\tau^{*}$ in the following -
one of the two layers (in particular, the smallest one) switches the components of its magnetization vector
so to be aligned with the other layer. The free energy corresponding to these solutions is shown again in panel \figpanel{a5} (dashed green line): it should be noticed how, at $\tau<\tau^{*}$, these solutions are thermodynamically unfavored,
i.e. their free energy is higher than the solution corresponding to
the same pattern index (except in the trivial case at $\tau=0$ where
the two Hopfield models are independent), so in this sense they are
metastable. 
To summarize, while varying $\tau$ there is first a (trivial) phase co-existence at $\tau=0$; then, the recall of two distinct patterns is locally stable, but unfavoured, until the state completely disappear at $\tau^{*}$. Such threshold values are computed numerically and plotted Figure  \ref{fig:sectionV} \figpanel{b} as functions of $\beta>1$ for different values of $\gamma$. The value of $\tau^{*}$ turns out to increase upon lowering the temperature for a fixed $\gamma$ and viceversa at fixed $\beta$ when $\gamma$ departs from $1$.

\section{Conclusions\label{sec:Conclusion}}
In this paper, we have presented a comprehensive and exhaustive statistical mechanics treatment of the Bidirectional Associative Memory (BAM).
We have characterized the phase diagram of the BAM, and in particular its operating regimes, i.e., the existence and extent of its retrieval phase, by both an analytic treatment in the asymptotic limit and numerical simulations, also by overcoming limitations of previous works.
We have paid particular attention to the effects of asymmetry between the sizes of the layers of the bipartite network and concluded that the asymmetry tends  to damage the retrieval capacities. 
We have also shown that the  BAM can be used as an efficient alternative to the Hopfield model from the point of view of the storage and retrieval properties of recurrent neural networks, in terms of the number of weights required to store a fixed set of patterns, and the efficiency of the reverberation mechanism explained in Section \ref{sec:numericalsim} using parallel dynamics, the latter allowing significant speedup of sampling mechanisms (particularly efficient on GPUs), compared to the standard Hopfield model. However, the price to be paid is, on the one hand, a lower network capacity compared to the (fully connected) Hopfield model and, on the other hand, a smaller basin of attraction for the smallest layer in the asymmetric BAM. 
In principle, this phenomenom could be further amplified by adding more and more layers: the resulting network might retrieve an overall smaller number of patterns than in the Hopfield case, but the efficiency of storage would be improved by the structure of the network. We leave this topic for future investigations.\\
Additionally, this work lays the foundation for the use of BAM as an energy-based generative model in unsupervised learning. An earlier attempt to use the retrieval properties of the Hopfield model in training RBMs was recently proposed~\citep{Pozas_Kerstjens_2021}, but such a training process is likely to introduce correlations between patterns when using real datasets. This is a situation we have not considered in this work and should be addressed in order to  properly characterize the model's phase diagram during learning. Although, for instance, the Hopfield model with correlated patterns has already been analyzed in the simple low-load regime ~\citep{agliari2013parallel}, we still lack a general theory that can be used in a context of learning.

\begin{acknowledgments}
G.C., A.D. and B.S. acknowledge financial support by the Comunidad de Madrid and the Complutense University of Madrid (Spain) through the Atracción de Talento programs (Refs. 2019-T1/TIC-13298 for A.D. and G.C. and 2019-T1/TIC-12776 for B.S.), the Banco Santander and the UCM (grant PR44/21-29937), and Ministerio de Econom\'{\i}a y Competitividad, Agencia Estatal de Investigaci\'on and Fondo Europeo de Desarrollo Regional (FEDER) (Spain and European Union) through the grant PID2021-125506NA-I00. AB acknowledges financial support by Ministero degli Affari Esteri e della Cooperazione Internazionale (Italy-Israel 
collaboration, BULBUL grant, CUP Project n. F85F21006230001). G.C. acknowledges useful discussions with Linda Albanese and Alberto Fachechi. All the authors acknowledge hospitality during the Alan Turing Institute's Theory and Methods Challenge Fortnights event ``Physics-informed machine learning''.
\end{acknowledgments}

\newpage
\appendix

\section{Interpolation Method\label{sec:Interpolation-Method}}
We now illustrate how to compute the BAM's quenched free energy using
the interpolation technique developed by Guerra \citep{guerra_course_2006,guerra_sumrules}.
We proceed by analogy with previous related works on spin-glass models
(see e.g. \citep{barra_how_2012} ), by defining the following quenched
intensive pressure at fixed size $L$
\begin{equation}
\mathcal{A}_{L}\left(\beta,\alpha,\gamma\right)=\frac{1}{L}\mathbb{\mathbb{E}}_{\boldsymbol{\xi},\bar{\boldsymbol{\xi}}}\log\sum_{\boldsymbol{\sigma},\bar{\boldsymbol{\sigma}}}\exp\left[\frac{\beta}{L}\sum_{\mu=1}^{K}\left(\sum_{i=1}^{N}\xi_{i}^{\mu}\sigma_{i}\right)\left(\sum_{j=1}^{{\N}}\bar{\xi}_{j}^{\mu}\bar{\sigma}_{j}\right)\right],\label{eq:quenched_p_starting_appendix}
\end{equation}
and we are interested in evaluating its thermodynamic limit, namely
\begin{equation}
\mathcal{A}\left(\beta,\alpha,\gamma\right)=\lim_{L\to\infty}\mathcal{A}_{L}\left(\beta,\alpha,\gamma\right).
\end{equation}
The above quantity is related to the usual Helmoltz free energy density
through $f=-\mathcal{A}\slash\beta$. Considering the choice of pattern components for the signal and noise term discussed in Section \ref{subsec:basicsetting} of the main text, the operator $\mathbb{\mathbb{E}}_{\boldsymbol{\xi},\bar{\boldsymbol{\xi}}}$ reads
\begin{align}
\mathbb{\mathbb{E}}_{\boldsymbol{\xi},\bar{\boldsymbol{\xi}}} & =\prod_{i=1}^{N}\prod_{j=1}^{{\N}}\left[\prod_{\mu=1}^{l}\mathbb{E}_{\xi_{i}^{\mu}\in\left\{ -1,1\right\} }\mathbb{E}_{\bar{\xi}_{j}^{\mu}\in\left\{ -1,1\right\} }\right]\left[\prod_{\mu=l+1}^{K}\mathbb{E}_{\xi_{i}^{\mu}\sim\mathcal{N}\left(0,1\right)}\mathbb{E}_{\bar{\xi}_{j}^{\mu}\sim\mathcal{N}\left(0,1\right)}\right].
\end{align}

The interpolation method developed
by Guerra allows to compute Eq. \eqref{eq:quenched_p_starting_appendix}
in the thermodynamic limit $L\to\infty$ when the extensive set of disordered quantities
are Gaussian i.i.d. random variables. The interpolating structure
of the quenched free energy is constructed in terms of an interpolating
real parameter $t\in\left[0,1\right]$, whose expression is shown
below \footnote{unless otherwise specified, the symbol $\sum_{\mu}$ represents a
summation over the noisy part of the pattern, i.e. $\sum_{\mu}\equiv\sum_{\mu=l+1}^{K}$}:

\begin{align}
\mathcal{A}_{L}\left(\beta,\alpha,\gamma,t\right)= &\, \frac{1}{L}\mathbb{\mathbb{E}}\log\sum_{\boldsymbol{\sigma},\bar{\boldsymbol{\sigma}}}\exp\left\{ t\frac{\beta}{L}\sum_{\mu=1}^{l}\left(\sum_{i}\xi_{i}^{\mu}\sigma_{i}\right)\left(\sum_{j}\bar{\xi}_{j}^{\mu}\bar{\sigma}_{j}\right)+\left(1-t\right)\beta\sum_{\mu=1}^{l}\left[\Psi_{\mu}\sum_{i}\xi_{i}^{\mu}\sigma_{i}+\bar{\Psi}_{\mu}\sum_{j}\bar{\xi}_{j}^{\mu}\bar{\sigma}_{j}\right]\right\} \times\nonumber \\
 & \times\int\prod_{\mu}dz_{\mu}z_{\mu}^{\dagger}\exp\left\{ -\beta\sum_{\mu}z_{\mu}z_{\mu}^{\dagger}+\beta\sqrt{\frac{t}{N}}\sum_{i,\mu}z_{\mu}\xi_{i}^{\mu}\sigma_{i}+\beta\sqrt{\frac{t}{\N}}\sum_{j,\mu}z_{\mu}^{\dagger}\bar{\xi}_{j}^{\mu}\bar{\sigma}_{j}+\right.\nonumber \\
 & +\sqrt{1-t}\left(c\sum_{i}\eta_{i}\sigma_{i}+\bar{c}\sum_{j}\bar{\eta}_{j}\bar{\sigma}_{j}\right)+\sqrt{1-t}\left(d\sum_{\mu}y_{\mu}z_{\mu}+\bar{d}\sum_{\mu}\bar{y}_{\mu}z_{\mu}^{\dagger}\right)\nonumber \\
 & \left.+\frac{\left(1-t\right)}{2}\sum_{\mu}\left(z_{\mu},z_{\mu}^{\dagger}\right)\boldsymbol{E}\left(z_{\mu},z_{\mu}^{\dagger}\right)^{T}\right\},\label{eq:guerra_interpolated_press}
\end{align}

where we have already split the signal and noise term in the starting Hamiltonian, as discussed in the main text (see Section \ref{subsec:basicsetting}).
In the above expression, an additional set of i.i.d. standard Gaussian variables
has been introduced (whose average is included in the expectation $\mathbb{E}$), namely
$\left\{ \eta_{i}\right\} _{i=1}^{N},\left\{ \eta_{j}\right\} _{j=1}^{{\N}},\left\{ y_{\mu},\bar{y}_{\mu}\right\} _{\mu=l+1}^{K}$.
The scalars $\left\{ c,\bar{c},d,\bar{d}\right\} $ and the entries
of the $2\times2$ matrix $\boldsymbol{E}$ are constant real values
yet to be fixed. It is trivial to check that the original intensive
pressure Eq. \eqref{eq:quenched_p_starting_appendix} is recovered
when $t=1$. On the other hand, at $t=0$ the interacting term in
the original Hamiltonian disappears, and the corresponding intensive
pressure is that of a factorized model over the spin components (on
both layers): the only difference here with the standard Hopfield
model is that the $t=0$ case is contains a series of $2-$body problems
in the pairs $\left(z_{\mu},z_{\mu}^{\dagger}\right)$, which is still
easy to handle. Notice that the two terms on the first line of Eq. \eqref{eq:guerra_interpolated_press}
- accounting for the signal (retrieval) part of the free energy -
are interpolated linearly with $t$, following the standard procedure
for the Curie-Weiss model \citep{barra_mean_2008}. All the others
are interpolated using the square root function as usually done for
spin-glasses. The key point behind such a method is that the quenched
pressure of the BAM, obtained by setting $t=1$, can be obtained exploiting
the calculus theorem:

\begin{equation}
\mathcal{A}_{L}\left(\beta,\alpha,\gamma,t=1\right)=\mathcal{A}_{L}\left(\beta,\alpha,\gamma,t=0\right)+\int_{0}^{1}dt'\frac{d\mathcal{A}_{L}\left(\beta,\alpha,\gamma,t'\right)}{dt'}\, . \label{eq:calculus_theoremm}
\end{equation}
Therefore, we just need to compute the initial (Cauchy) condition
at $t=0$ and the $t-$derivative of the interpolating quenched pressure.
\subsection{Derivative}
We now evaluate the $t-$derivative of Eq. \eqref{eq:guerra_interpolated_press}.
For simplicity, let us call the interpolated exponent of \eqref{eq:guerra_interpolated_press}
as $-H\left(t\right)$: in the following, we will indicate with the
letter $\omega\left(\mathcal{O}\right)$ the expectation value of
a certain observable $\mathcal{O}$ w.r.t. the Boltzmann measure for
a fixed realization of the quenched disorder:
\begin{equation}
\omega\left(\mathcal{O}\right)=\frac{\sum_{\boldsymbol{\sigma},\bar{\boldsymbol{\sigma}}}\int\prod_{\mu}dz_{\mu}z_{\mu}^{\dagger}\mathcal{O}\exp\left[-H\left(t\right)\right]}{\sum_{\boldsymbol{\sigma},\bar{\boldsymbol{\sigma}}}\int\prod_{\mu}dz_{\mu}z_{\mu}^{\dagger}\exp\left[-H\left(t\right)\right]},\label{eq:boltzmann_exp_omega}
\end{equation}
and wih $\langle\mathcal{O}\rangle=\mathbb{E}\omega\left(\mathcal{O}\right)$: from now on, the operator $\mathbb{E}$ denotes the expectation value over all the possible disorder components appearing in the quenched pressure.
The total $t-$derivative of \eqref{eq:guerra_interpolated_press}
reads:
\begin{align}
\frac{d\mathcal{A}}{dt} =&+ \frac{\beta}{L^{2}}\sum_{\mu=1}^{l}\mathbb{E}\omega\left[\left(\sum_{i}\xi_{i}^{\mu}\sigma_{i}\right)\left(\sum_{j}\bar{\xi}_{j}^{\mu}\bar{\sigma}_{j}\right)\right]-\frac{\beta}{L}\sum_{\mu=1}^{l}\mathbb{E}\omega\left[\Psi_{\mu}\sum_{i}\xi_{i}^{\mu}\sigma_{i}+\bar{\Psi}_{\mu}\sum_{j}\bar{\xi}_{j}^{\mu}\bar{\sigma}_{j}\right]+\nonumber \\
 & +\frac{\beta}{2L\sqrt{N}\sqrt{t}}\sum_{i,\mu}\mathbb{E}\omega\left(z_{\mu}\xi_{i}^{\mu}\sigma_{i}\right)+\frac{\beta}{2L\sqrt{{\N}}\sqrt{t}}\sum_{j,\mu}\mathbb{E}\omega\left(z_{\mu}^{\dagger}\bar{\xi}_{j}^{\mu}\bar{\sigma}_{j}\right)-\frac{c}{2L\sqrt{1-t}}\sum_{i}\mathbb{E}\omega\left(\eta_{i}\sigma_{i}\right)-\frac{\bar{c}}{2L\sqrt{1-t}}\sum_{j}\mathbb{E}\omega\left(\bar{\eta}_{j}\bar{\sigma}_{j}\right)\nonumber \\
 & -\frac{d}{2L\sqrt{1-t}}\sum_{\mu}\mathbb{E}\omega\left(y_{\mu}z_{\mu}\right)-\frac{\bar{d}}{2L\sqrt{1-t}}\sum_{\mu}^{K}\mathbb{E}\omega\left(\bar{y}_{\mu}z_{\mu}^{\dagger}\right)-\frac{1}{2L}\sum_{\mu}^{K}\mathbb{E}\omega\left[\left(z_{\mu},z_{\mu}^{\dagger}\right)\boldsymbol{E}\left(z_{\mu},z_{\mu}^{\dagger}\right)^{T}\right].\label{eq:stream_first}
\end{align}
All the terms explicitly depending on Gaussian quenched disordered
can be further simplified by exploiting Wick's Theorem:
\begin{equation}
\mathbb{E}\left[x f\left(x\right)\right]=\mathbb{E}\left[\partial_{x}f\left(x\right)\right],\label{eq:Wickstheorem}
\end{equation}
where $x\sim\mathcal{N}\left(0,1\right)$ is a standard normal random
variable. After applying Eq. \eqref{eq:Wickstheorem} to the above
derivative, we get:
\begin{align}
\frac{d\mathcal{A}}{dt} =&\; \frac{\beta}{L^{2}}\sum_{\mu=1}^{l}\mathbb{E}\omega\left[\left(\sum_{i}\xi_{i}^{\mu}\sigma_{i}\right)\left(\sum_{j}\bar{\xi}_{j}^{\mu}\bar{\sigma}_{j}\right)\right]-\frac{\beta}{L}\sum_{\mu=1}^{l}\mathbb{E}\omega\left[\Psi_{\mu}\sum_{i}\xi_{i}^{\mu}\sigma_{i}+\bar{\Psi}_{\mu}\sum_{j}\bar{\xi}_{j}^{\mu}\bar{\sigma}_{j}\right]+\nonumber \\
 & +\frac{\beta^{2}}{2LN}\sum_{i,\mu}\mathbb{E}\left[\omega\left(z_{\mu}^{2}\right)-\omega^{2}\left(z_{\mu}\sigma_{i}\right)\right]+\frac{\beta^{2}}{2L{\N}}\sum_{j,\mu}\mathbb{E}\left[\omega\left(z_{\mu}^{\dagger2}\right)-\omega^{2}\left(z_{\mu}^{\dagger}\bar{\sigma}_{j}\right)\right]\nonumber \\
 & -\frac{c^{2}}{2L}\sum_{i}\mathbb{E}\left[1-\omega^{2}\left(\sigma_{i}\right)\right]-\frac{\bar{c}^{2}}{2L}\sum_{j}\mathbb{E}\left[1-\omega^{2}\left(\bar{\sigma}_{j}\right)\right] \nonumber \\
 & -\frac{d^{2}}{2L}\sum_{\mu}\mathbb{E}\left[\omega\left(z_{\mu}^{2}\right)-\omega^{2}\left(z_{\mu}\right)\right]-\frac{\bar{d}\,^{2}}{2L}\sum_{\mu}\mathbb{E}\left[\omega\left(z_{\mu}^{\dagger2}\right)-\omega^{2}\left(z_{\mu}^{\dagger}\right)\right]-\frac{1}{2L}\sum_{\mu}^{K}\mathbb{E}\omega\left[\left(z_{\mu},z_{\mu}^{\dagger}\right)\boldsymbol{E}\left(z_{\mu},z_{\mu}^{\dagger}\right)^{T}\right].\label{eq:stream_afterwick}
\end{align}
Now, introducing the order parameters using the definitions given
in the main text, namely Eqs. \eqref{eq:mattis_both}-\eqref{eq:overlapdef_both}-\eqref{eq:overlapP_def_both}
we can further simplify the above expression as follows:
\begin{align}
\frac{d\mathcal{A}}{dt} =&\; \beta\sum_{\mu=1}^{l}\left[\left\langle m_{\mu}\bar{m}_{\mu}\right\rangle -\gamma\Psi_{\mu}\left\langle m_{\mu}\right\rangle -\bar{\gamma}\bar{\Psi}_{\mu}\left\langle \bar{m}_{\mu}\right\rangle \right]+\\
 & -\frac{\alpha\beta^{2}}{2}\left\langle q_{\repidx{}}p_{\repidx{}}\right\rangle -\frac{\alpha\beta^{2}}{2}\left\langle \bar{q}_{\repidx{}}\bar{p}_{\repidx{}}\right\rangle -\frac{\gamma c^{2}}{2}\left(1-\left\langle q_{\repidx{}}\right\rangle \right)-\frac{\bar{\gamma}\bar{c}^{2}}{2}\left(1-\left\langle \bar{q}_{\repidx{}}\right\rangle \right)\\
 & -\frac{\alpha d^{2}}{2}\left\langle p_{\repidx{}}\right\rangle -\frac{\alpha\bar{d}\,^{2}}{2}\left\langle \bar{p}_{\repidx{}}\right\rangle -\frac{1}{2L}\sum_{\mu=l+1}^{K}\left\langle \left(z_{\mu},z_{\mu}^{\dagger}\right)\tilde{\boldsymbol{E}}\left(z_{\mu},z_{\mu}^{\dagger}\right)^{T}\right\rangle ,
\end{align}
where
\begin{equation}
\tilde{\boldsymbol{E}}=\boldsymbol{E}-\begin{bmatrix}\beta^{2}-d^{2} & 0\\
0 & \beta^{2}-\bar{d}\,^{2}
\end{bmatrix}.\label{eq:matrix_Etilde}
\end{equation}
The main idea now is to fix the scalar parameters $ c,\bar{c},d,\bar{d},\left\{ E_{ij}\right\} $
in such a way to explicitly show the fluctuations of the order parameters.
First notice that the above expression can be rewritten as
\begin{align}
\frac{d\mathcal{A}}{dt}  =& \; +\beta\sum_{\mu=1}^{l}\left[\left\langle m_{\mu}\bar{m}_{\mu}\right\rangle -\gamma\Psi_{\mu}\left\langle m_{\mu}\right\rangle -\bar{\gamma}\bar{\Psi}_{\mu}\left\langle \bar{m}_{\mu}\right\rangle \right]\nonumber \\
 & -\frac{\alpha\beta^{2}}{2}\left[\left\langle q_{\repidx{}}p_{\repidx{}}\right\rangle -\frac{\gamma c^{2}}{\alpha\beta^{2}}\left\langle q_{\repidx{}}\right\rangle -\frac{d^{2}}{\beta^{2}}\left\langle p_{\repidx{}}\right\rangle \right]-\frac{\gamma c^{2}}{2}\nonumber \\
 & -\frac{\alpha\beta^{2}}{2}\left[\left\langle \bar{q}_{\repidx{}}\bar{p}_{\repidx{}}\right\rangle -\frac{\bar{\gamma}\bar{c}^{2}}{\alpha\beta^{2}}\left\langle \bar{q}_{\repidx{}}\right\rangle -\frac{\bar{d}\,^{2}}{\beta^{2}}\left\langle \bar{p}_{\repidx{}}\right\rangle \right]-\frac{\gamma\bar{c}^{2}}{2}\nonumber \\
 & -\frac{1}{2L}\sum_{\mu}\left\langle \left(z_{\mu},z_{\mu}^{\dagger}\right)\tilde{\boldsymbol{E}}\left(z_{\mu},z_{\mu}^{\dagger}\right)^{T}\right\rangle. \label{eq:total_stream_afteroverlaps-1}
\end{align}
Finally, by fixing the scalar parameters as follows
\begin{equation}
\begin{array}{ccccc}
\Psi_{\mu}=\bar{\gamma}\left\langle \bar{m}_{\mu}\right\rangle  &  &  &  & \bar{\Psi}_{\mu}=\gamma\left\langle m_{\mu}\right\rangle \\
\\
c=\beta\sqrt{\bar{\gamma}\alpha\left\langle p_{\repidx{}}\right\rangle } &  &  &  & \bar{c}=\beta\sqrt{\gamma\alpha\left\langle \bar{p}_{\repidx{}}\right\rangle }\\
\\
d=\beta\sqrt{\left\langle q_{\repidx{}}\right\rangle } &  &  &  & \bar{d}=\beta\sqrt{\left\langle \bar{q}_{\repidx{}}\right\rangle }\\
\\
E_{11}=\beta^{2}-d^{2} &  &  &  & E_{22}=\beta^{2}-\bar{d}\,^{2}
\end{array}\label{eq:fixing_parameters_RS}
\end{equation}
 and $E_{12}=E_{21}=0$ we can simplify the total stream as:
\begin{align}
\frac{d\mathcal{A}}{dt} = & \; \beta\sum_{\mu=1}^{l}\left\langle \left(m_{\mu}-\left\langle m_{\mu}\right\rangle \right)\left(\bar{m}_{\mu}-\left\langle \bar{m}_{\mu}\right\rangle \right)\right\rangle -\beta\sum_{\mu=1}^{l}\left\langle m_{\mu}\bar{m}_{\mu}\right\rangle \nonumber \\
 &-\frac{\alpha\beta^{2}}{2}\left\langle \left(q_{\repidx{}}-\left\langle q_{\repidx{}}\right\rangle \right)\left(p_{\repidx{}}-\left\langle p_{\repidx{}}\right\rangle \right)\right\rangle -\frac{\alpha\beta^{2}}{2}\left\langle p_{\repidx{}}\right\rangle \left(1-\left\langle q_{\repidx{}}\right\rangle \right)\nonumber \\
 & -\frac{\alpha\beta^{2}}{2}\left\langle \left(\bar{q}_{\repidx{}}-\left\langle \bar{q}_{\repidx{}}\right\rangle \right)\left(\bar{p}_{\repidx{}}-\left\langle \bar{p}_{\repidx{}}\right\rangle \right)\right\rangle -\frac{\alpha\beta^{2}}{2}\left\langle \bar{p}_{\repidx{}}\right\rangle \left(1-\left\langle \bar{q}_{\repidx{}}\right\rangle \right).\label{eq:total_stream_aftersimplification}
\end{align}

\subsubsection{Replica symmetric solution}

Under the replica symmetric (RS) approximation, all the order parameters
concentrate around a fixed value (to be determined self-consistently at any $\alpha,\beta,\gamma$) in the thermodynamic limit and do
not fluctuate, namely
\[
\begin{array}{ccccccc}
m_{\mu}\stackrel{\textsc{RS}}{\to}M_{\mu} &  &  & \langle q_{\repidx{}}\rangle\stackrel{\textsc{RS}}{\to}Q &  &  & \langle p_{\repidx{}}\rangle\stackrel{\textsc{RS}}{\to}P\\
\\
\bar{m}_{\mu}\stackrel{\textsc{RS}}{\to}\bar{M}_{\mu} &  &  & \langle\bar{q}_{\repidx{}}\rangle\stackrel{\textsc{RS}}{\to}\bar{Q} &  &  & \langle\bar{p}_{\repidx{}}\rangle\stackrel{\textsc{RS}}{\to}\bar{P}
\end{array}
\]
With this assumption, all the fluctuation terms in \eqref{eq:total_stream_aftersimplification}
disappear, and the final expression of the time derivative for the
interpolating quenched pressure simplifies to
\begin{equation}
\frac{d\mathcal{A}}{dt}=-\beta\sum_{\mu=1}^{l}M_{\mu}\bar{M}_{\mu}-\frac{\alpha\beta^{2}}{2}P\left(1-Q\right)-\frac{\alpha\beta^{2}}{2}\bar{P}\left(1-\bar{Q}\right)\label{eq:RS_stream}
\end{equation}

\subsection{Initial condition}

The last missing piece to evaluate \eqref{eq:total_stream_aftersimplification}
is the initial (Cauchy) condition at $t=0$. As already discussed,
such a computation is trivial as the interacting part disappears and
we are left with a series of one-body terms w.r.t. to all the spin
variables, plu set of 2-body problems in the pairs $\left(z_{\mu},z_{\mu}^{\dagger}\right)$.
The quenched pressure evaluated at $t=0$ reads:
\begin{multline}
\mathcal{A}_{L}\left(\beta,\alpha,\gamma,t=0\right)=\frac{1}{L}\mathbb{\mathbb{E}}\log\sum_{\boldsymbol{\sigma},\bar{\boldsymbol{\sigma}}}\exp\left[\beta\sum_{\mu=1}^{l}\left(\Psi_{\mu}\sum_{i}\xi_{i}^{\mu}\sigma_{i}+\bar{\Psi}_{\mu}\sum_{j}\bar{\xi}_{j}^{\mu}\bar{\sigma}_{j}\right)+c\sum_{i}\eta_{i}\sigma_{i}+\bar{c}\sum_{j}\bar{\eta}_{j}\bar{\sigma}_{j}\right]\times\\
\times\prod_{\mu}\int dz_{\mu}z_{\mu}^{\dagger}\exp\left[-\beta z_{\mu}z_{\mu}^{\dagger}+dy_{\mu}z_{\mu}+\bar{d}\bar{y}_{\mu}z_{\mu}^{\dagger}+\frac{1}{2}\left(z_{\mu},z_{\mu}^{\dagger}\right)\boldsymbol{E}\left(z_{\mu},z_{\mu}^{\dagger}\right)^{T}\right].\label{eq:A0_start}
\end{multline}
The traces over $\boldsymbol{\sigma},\bar{\boldsymbol{\sigma}}$ can
be easily computed since everything is factorized.
Moreover, each integral in the second line (they are all equal) can
be transformed into a Gaussian integral over a pair of real variables
$u_{\mu},v_{\mu}$ by exploiting $z_{\mu}=u_{\mu}+iv_{\mu}$ and $z_{\mu}^{\dagger}=u_{\mu}-iv_{\mu}$.
After some calculations, we get:
\begin{align}
\mathcal{A}_{L}\left(\beta,\alpha,\gamma,t=0\right) =&\; \gamma\mathbb{E}_{\eta,\boldsymbol{\xi}}\log2\text{cosh}\left[c\eta+\beta\sum_{\mu=1}^{l}\Psi_{\mu}\xi^{\mu}\right]+\bar{\gamma}\mathbb{E}_{\bar{\eta},\bar{\boldsymbol{\xi}}}\log2\text{cosh}\left[\bar{c}\bar{\eta}+\beta\sum_{\mu=1}^{l}\bar{\Psi}_{\mu}\bar{\xi}^{\mu}\right]+\nonumber \\
 & -\frac{\alpha}{2}\log\text{det}\boldsymbol{K}+\frac{\alpha}{2}\mathbb{E}_{y,\bar{y}}\left[\left(dy+\bar{d}\bar{y},idy-i\bar{d}\bar{y}\right)\boldsymbol{K}^{-1}\left(dy+\bar{d}\bar{y},idy-i\bar{d}\bar{y}\right)^{T}\right],\label{eq:A0_after_traces}
\end{align}
where the matrix \textbf{$\boldsymbol{K}$} is given by
\begin{equation}
\boldsymbol{K}=\begin{bmatrix}2\beta-\left(E_{11}+E_{22}\right) & i\left(E_{22}-E_{11}\right)\\
i\left(E_{22}-E_{11}\right) & 2\beta+\left(E_{11}+E_{22}\right)
\end{bmatrix}.\label{eq:matrix_K}
\end{equation}
The last term in the second line of Eq. \eqref{eq:A0_after_traces}
can be further simplified by explicitly computing the average over
$y,\bar{y}$, being two i.i.d standard Gaussian variables. After substituting
all the scalar parameters as fixed by \eqref{eq:fixing_parameters_RS}
within the RS ansatz, the final expression for the initial condition
on the quenched intensive pressure reads:
\begin{align}
\mathcal{A}_{L}\left(\beta,\alpha,\gamma,t=0\right) =& \; \gamma\mathbb{E}_{\eta,\boldsymbol{\xi}}\log2\text{cosh}\left[\beta\sqrt{\bar{\gamma}\alpha P}\eta+\beta\bar{\gamma}\sum_{\mu}\bar{M}_{\mu}\xi^{\mu}\right]+\bar{\gamma}\mathbb{E}_{\bar{\eta},\bar{\boldsymbol{\xi}}}\log2\text{cosh}\left[\beta\sqrt{\gamma\alpha\bar{P}}\bar{\eta}+\beta\gamma\sum_{\mu}M_{\mu}\bar{\xi}^{\mu}\right]+\nonumber \\
 & -\frac{\alpha}{2}\log\left[1-\beta^{2}\left(1-Q\right)\left(1-\bar{Q}\right)\right]+\frac{\alpha\beta^{2}}{2}\frac{\left[Q\left(1-\bar{Q}\right)+\bar{Q}\left(1-Q\right)\right]}{1-\beta^{2}\left(1-Q\right)\left(1-\bar{Q}\right)}.\label{eq:initial_condition_final}
\end{align}
Finally, putting together Eqs. \eqref{eq:initial_condition_final} and
\eqref{eq:RS_stream}, and using the calculus theorem \eqref{eq:calculus_theoremm}
we get the expression for the RS quenched intensive pressure:

\begin{align}
\mathcal{A}\left(\beta,\alpha,\gamma\right) =& -\beta\sum_{\mu=1}^{l}M_{\mu}\bar{M}_{\mu}-\frac{\alpha\beta^{2}}{2}P\left(1-Q\right)-\frac{\alpha\beta^{2}}{2}\bar{P}\left(1-\bar{Q}\right)+\nonumber \\
 & +\gamma\mathbb{E}_{\eta,\boldsymbol{\xi}}\log2\text{cosh}\left[\beta\sqrt{\bar{\gamma}\alpha P}\eta+\beta\bar{\gamma}\sum_{\mu}\bar{M}_{\mu}\xi^{\mu}\right]+\bar{\gamma}\mathbb{E}_{\bar{\eta},\bar{\boldsymbol{\xi}}}\log2\text{cosh}\left[\beta\sqrt{\gamma\alpha\bar{P}}\bar{\eta}+\beta\gamma\sum_{\mu}M_{\mu}\bar{\xi}^{\mu}\right]+\nonumber \\
 & -\frac{\alpha}{2}\log\left[1-\beta^{2}\left(1-Q\right)\left(1-\bar{Q}\right)\right]+\frac{\alpha\beta^{2}}{2}\frac{\left[Q\left(1-\bar{Q}\right)+\bar{Q}\left(1-Q\right)\right]}{1-\beta^{2}\left(1-Q\right)\left(1-\bar{Q}\right)}\nonumber \\
 \equiv & -\beta f\left(\beta,\alpha,\gamma\right)
 \label{eq:total_RS_quenchedpressure}
\end{align}
with $f\left(\beta,\alpha,\gamma\right)$ being the RS free energy expression in the main text (Eq. \eqref{eq:RS_freeenergy}). 



\section{Additional numerical results \label{sec:appendix_seqdyn}}

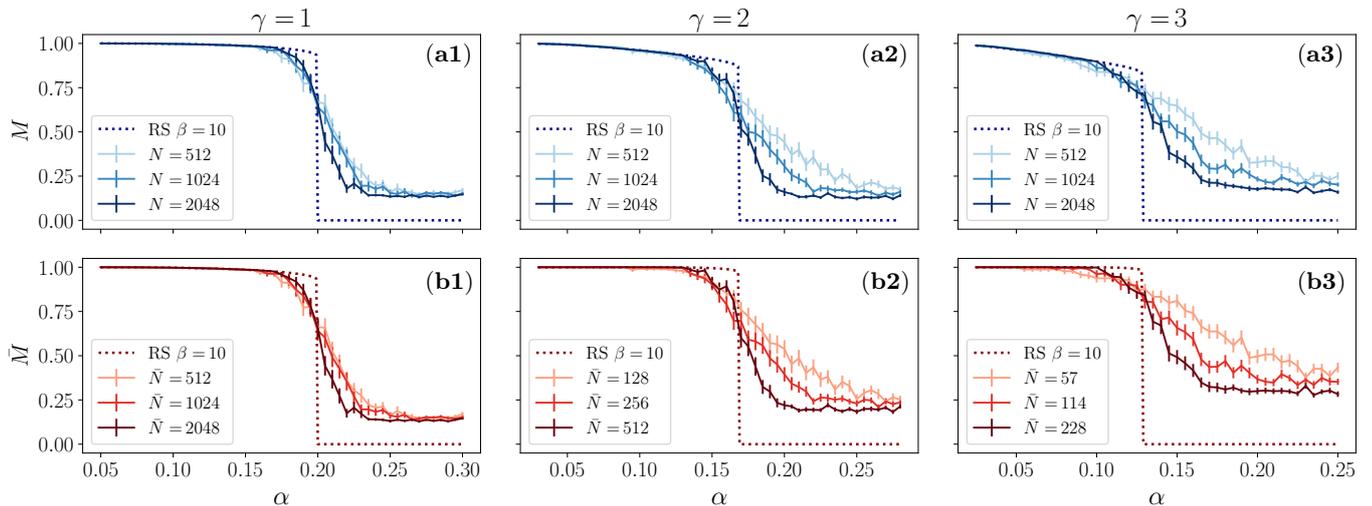
\begin{figure}
\begin{overpic}[width=\textwidth]{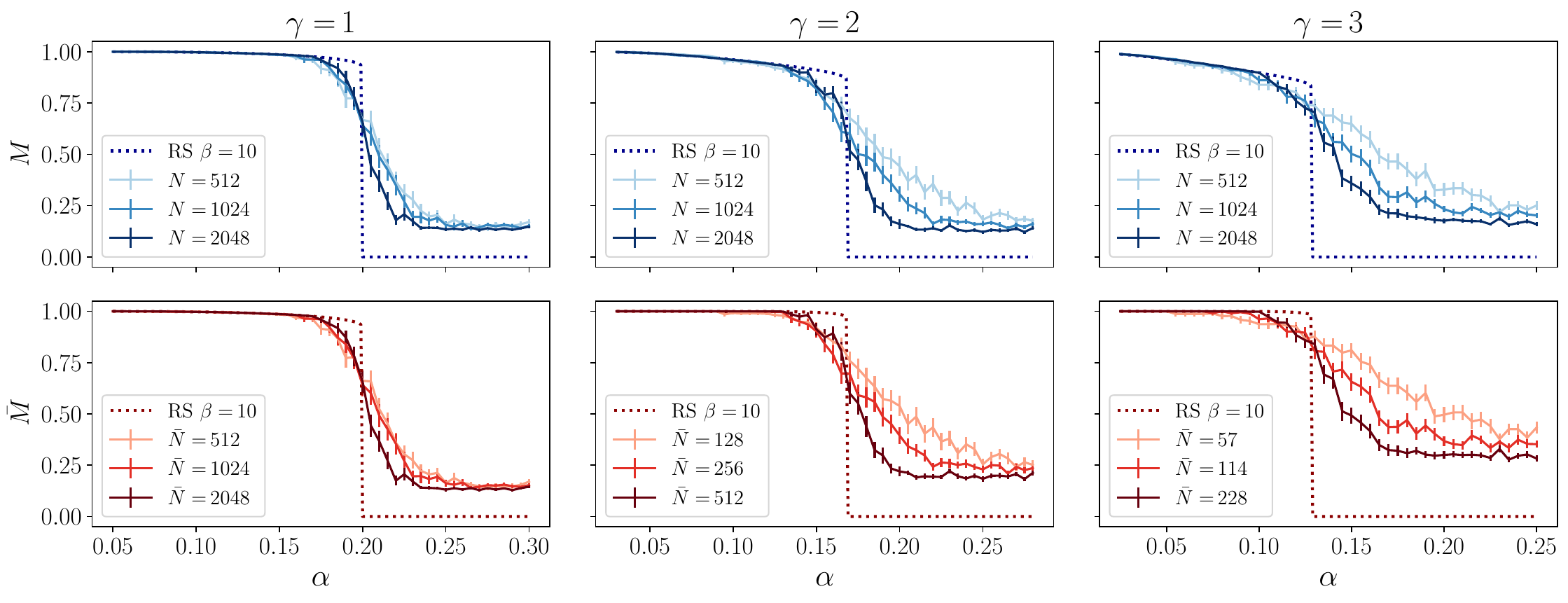}
\put(31,33.8){{\figpanel{a1}}}
\put(63,33.8){{\figpanel{a2}}}
\put(95,33.8){{\figpanel{a3}}}
\put(31,17){{\figpanel{b1}}}
\put(63,17){{\figpanel{b2}}}
\put(95,17.){{\figpanel{b3}}}
\end{overpic}

\caption{Numerical estimation of the retrieval/spin-glass phase transition. Simulations are performed using a random sequential dynamics at $\beta=10$ with $N_{\text{steps}} = 10^4$ for $3$ values of $\gamma$. \figpanel{a1}-\figpanel{b1}: $\gamma=1$; \figpanel{a2}-\figpanel{b2}: $\gamma=2$;
\figpanel{a3}-\figpanel{b3}: $\gamma=3$;
Each panel shows the final Mattis magnetization over each layer w.r.t. to the pattern pair to be retrieved. In all cases, the dynamics start from the pattern pair to be retrieved with a small noise $\varepsilon_{1}=\varepsilon_{2}=0.05$ on both layers. Results are plotted as a function of $\alpha$ for three different system sizes (shown in the caption) and averaged over $50$ different realizations of the disorder.
Dashed lines in each panel denote the RS saddle point solution at $\beta=10$ (i.e. a horizontal cut in the phase diagrams of Fig. \ref{fig:Phasediagrams_all}).
\label{fig:storage_seqdynamics}}
    
\end{figure}

We report additional results on the numerical estimation of the retrieval/ spin-glass spinodal point at low temperatures, by using a random sequential dynamics (whose update rules are given by Eqs. \eqref{eq:dyn_beta_sigma_i_tp1}-\eqref{eq:dyn_beta_sigmabar_j_tp1} in Section \ref{sec:numericalsim}).
In all the simulations shown in Figure \ref{fig:storage_seqdynamics}, the dynamics starts with an initial configuration having noises $\varepsilon_{1}=\varepsilon_{2}=0.05$ on both layers. 
At each step, the dynamics is carried out by selecting one neuron among the union set of the two layers and running Eqs. \eqref{eq:dyn_beta_sigma_i_tp1} or \eqref{eq:dyn_beta_sigmabar_j_tp1},  depending on which layer the selected neuron belongs to; this procedure is repeated for $(N+{\N}) \times N_{\text{steps}}$ single neuron updates, so that -on average- each neuron is updated $N_{\text{steps}}$ times.
In Figure \ref{fig:storage_seqdynamics} the system sizes are smaller than the ones shown in Figure \ref{fig:storage_pardynamics} of the main text: the latter would require an unreasonable amount of time using a random sequential dynamics.
In general, despite the curve are less sharp than in Fig. \ref{fig:storage_pardynamics} of the main text (again, because of the smaller sizes), we still find an overall good agreement between the numerical prediction of the transition w.r.t. the MF theory.
We have numerically verified that these results remain qualitatively unchanged when the noise on both layers is increased up to $\varepsilon_{1}= \varepsilon_{2} \approx 0.3$: in the strongly asymmetric case, however, using this type of dynamics might affect the basin of attraction of the two layers in different ways w.r.t. what discussed in Section \ref{sec:numericalsim} of the main text.
As a final remark, note that the initial configuration on both layers must be constructed in such a way that the starting Mattis magnetizations have the same sign: otherwise, retrieval is lost and and both Mattis magnetizations will tend to $0$ (even at $\alpha \ll \alpha_c(\gamma)$), which is a clear consequence of the saddle type of the BAM Hamiltonian \eqref{eq:hamiltonian_starting}.

\section{Equivalence between BAM and coupled RBMs \label{sec:Equivalence_coupledRBMS}}
\begin{figure}
\begin{overpic}
[width=0.9\textwidth]{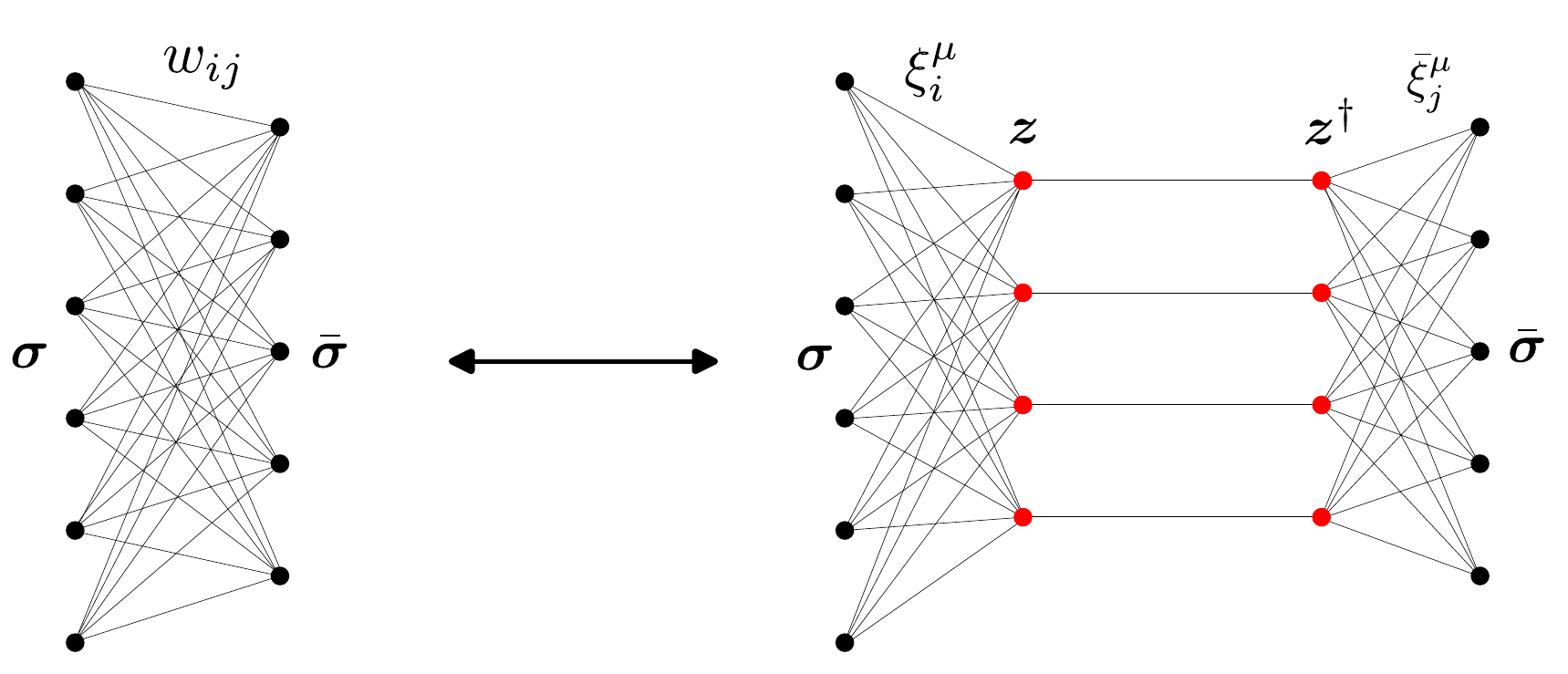}
\put(0,39){{\figpanel{a}}}
\put(48,39){{\figpanel{b}}}
\end{overpic}

\caption{Schematic representation of the equivalence between the BAM \figpanel{a} and two coupled RBMs \figpanel{b}. In \figpanel{a}, each link is a synaptic connection in the BAM (constructed using the Hebb rule through Eq. \eqref{eq:coupling_matrix}). In \figpanel{b}, the left-most and right-most interactions correspond to the patterns components $\xi_{i}^{\mu}$ and $\bar{\xi}_{j}^{\mu}$ respectively. Red dots represent hidden nodes acting on each layer, whose number is equal to the number of patterns $K$ in the BAM. \label{fig:Equivalence-between-BAM}}
\end{figure}

Recent works \citep{barra_equivalence_2012} have shown how the Hopfield model can be mapped into a binary-Gaussian RBM whose weights are directly linked to the pattern components stored in the Hopfield counterpart. Such an analogy provides an efficient way to determine the operating regimes of this specific type of artificial network for unsupervised learning: speficically, it allows on one hand to simulate more efficiently the retrieval properties of the Hopfield model  \citep{barra_equivalence_2012} and, on the other hand, to describe the functioning regimes of the artificial (RBM) counterpart in terms of the ratio between the number of hidden and visible variables, which further allow for efficient and computationally cheap pre-training procedures \citep{Leonelli-NN2021}. 
In principle, such an analogy can be extended to the BAM, at least from a structural point of view: indeed, considering the integral transformation \eqref{eq:integral_transform_complex_generic} used to decouple the interacting term in the BAM Hamiltonian \eqref{eq:hamiltonian_starting}, it is possible to interpret the right hand side of Eq. \eqref{eq:quenched_p_appendix_after_decouplings} as the partition function of a $4-$partite system similar to two coupled RBMs. This is shown in Figure \ref{fig:Equivalence-between-BAM}: the two RBMs on \figpanel{b} have each one a binary visible layer (with sizes $N$ and $\N$ respectively); the two hidden layers are encoded in the vectors  $\boldsymbol{z},\boldsymbol{z}^{\dagger}$: each of them has a size equal to the number of patterns in the BAM ($K$), and for each $\mu\in \{1, \ldots,K \}$  the hidden nodes $z_{\mu}, z^{\dagger}_{\mu}$ are complex conjugates, interacting through a fixed potential.

However, even if the analogy seems to be consistent from a structural point of view, it is not clear how to exploit it to perform learning in the resulting artificial network depicted in Figure \ref{fig:Equivalence-between-BAM} \figpanel{b}: for instance, gradient ascent on the pattern components would require sampling from a probability distribution defined on complex variables. We leave this issue for future investigation.

\end{document}